# Simulating graphene impurities using the worm algorithm

**Marcin Szyniszewski, BSc,** Major Physics Project
Supervisor: **Evgeni Burovski, PhD**
Department of Physics, Lancaster University

**Abstract**

The two-dimensional Ising model is studied by performing computer simulations with one of the Monte Carlo algorithms – the worm algorithm. The critical temperature $T_C$ of the phase transition is calculated by the usage of the critical exponents and the results are compared to the analytical result, giving a very high accuracy.

We also show that the magnetic ordering of impurities distributed on a graphene sheet is possible, by simulating the properly constructed model using the worm algorithm. The value of $T_C$ is estimated. Furthermore, the dependence of $T_C$ on the interaction constants is explored. We outline how one can proceed in investigating this relation in the future.

# Table of contents

# Introduction

Graphene is thought to be the material of the future. But to see the possible applications of graphene, we firstly think about the problems with electronics, that we are facing. Nowadays, electronic devices are important part of everyone's life. In making them more efficient, we encounter a scaling problem – how many transistors can we place on the unit area of integrated circuit. This dependence against time is approximated by the Moore's Law [1]. Of course, there is a limit, where we hit the atomic sizes boundary, and this is the main cause of looking for new materials or techniques that will help us proceed with scaling down the circuits in the future.

Besides, most electronic devices are produced with substances hard to obtain, what increases the costs of buying them. There is also a difficulty of finding materials, that are very durable, but also very flexible. It is, in addition, almost impossible to find such objects, that will have very good electronic properties. All these problems might disappear someday because of graphene. It was discovered only recently, in 2004, and yet in 2010 Nobel Prize was given "for ground-breaking experiments regarding [graphene]" [2].

First of all, graphene is a very cheap material, as it can be produced from graphite, that the reader can find for example in his pencil. Secondly, graphene could be the key to bringing spintronics to everyday use. This technology would not depend on the charge of electron, but on its spin, thus opening many new possibilities. By introducing this new technology, we might be able to solve the scaling problem for some time. This application and other unique electronic properties of graphene give us hope, that maybe in the near future there will be a turn in a history of electronic devices.

The scientists still work on the process to manufacture graphene on a global scale [3]. Therefore, before it is possible, the physical properties of this material are thoroughly investigated. Interesting topic is the existence of impurities on the graphene sheet, that can significantly modify the electric properties of graphene [4] [5] [6]. Those atoms can be deliberately put on the graphene lattice or they can be the residue after the fabrication process. If the impurities are of magnetic nature, there is a possibility of magnetic ordering, which has not been investigated before. Thus, the phenomena, that is of main interest of this project, is the phase transition between the low-temperature ordered phase and the high-temperature disordered phase. Also, we investigate the behaviour of physical properties near the transition point.

There are many current experimental studies involving impurities. But with today's technology, we are facing a possibility to simulate graphene with impurities by using computers. The obtained results can be used later for setting up further experimental work in this area. Clearly, to simulate the system and to study it, the appropriate algorithm with good performance is needed. This is provided by the class of simulation methods, called Monte Carlo methods. The particular algorithm was selected, namely the worm algorithm.

To get familiar with this algorithm, we have firstly used it to simulate the two-dimensional Ising model. It is the simplest statistical model that shows the phase transition and thus, it is preferable to see how our designated algorithm will work in this case and if it is going to be efficient. Subsequently, the graphene impurities have been simulated. The aim of this project was to study the magnetic ordering of these impurities. One may suspect that the impurities display the phase transition and in fact, as we will show in this project, this is indeed the case. In particular, we will be interested in calculating the critical temperature.

Our work will be dealing with one particular arrangement of the interaction constants (i.e. the variables that characterize the interactions between the impurities). Other case has been already done before, and we will unite those two results to sketch the possible dependence between the critical temperature and the interaction constants.

The outline of this work is as follows:

In **Chapter 1** the reader is introduced to the background materials that are essential to understand this project. Firstly, there is an overview of the graphene, where we present the lattice of graphene, its band structure and, most importantly, the problem of impurities on the graphene lattice. Afterwards, we summarize the basic concepts of statistical physics, that will be necessary in our work, as well as the phase transition topic. There is also a review of the one- and two-dimensional Ising model.

**Chapter 2** outlines the Monte Carlo methods. We present the simulations of these theoretical methods and Metropolis update algorithm. Furthermore, we present the worm algorithm on a simple Ising model. The results of using the worm algorithm to simulate the Ising model are shown in **Chapter 3**. There is also an analysis of the results that ends with estimation of the critical temperature for the two-dimensional Ising model.

**Chapter 4** presents how one can build a model of graphene impurities by showing simplification steps. Then we use the worm algorithm to simulate this model. We state our results and the analysis of the results to estimate the temperature of the phase transition and the critical exponents.

In **Chapter 5** we discuss the conclusions that can be derived from out results and sketch possible relation between critical temperature and interaction constant.

# Chapter 1
# Background materials

## 1.1 Overview of the graphene

### 1.1.1 Description of the graphene and its lattice, band structure

Graphene can be defined as a single carbon layer of graphite structure, and this is the main reason of its name [7] [8]. Yet this definition is not independently formulated, as it uses the word "graphite". Thus, International Union of Pure and Applied Chemistry (IUPAC) suggests to use this kind of definition for **graphene** [9]:

> *Flat monolayer of carbon atoms forming a two-dimensional tightly packed honeycomb lattice.*

In fact, graphene is a building block for all different types of graphite structures, such as fullerenes, nanotubes or three-dimensional graphite and thus can be called "mother of all graphitic structures" [9].

*Graphene lattice*, as said before, is a honeycomb lattice (see Figure 1). What is very important, is that it consists two triangular sublattices (one marked with blue, other one with green). Certain properties of graphene structure elements are bounded with those two sublattices.

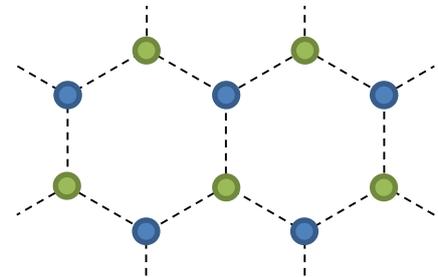

Figure 1. The graphene lattice with two sublattices.

The *primitive cell* of graphene, shown on Figure 2a contains two carbon atoms, each belonging to different sublattice. Of course, the choice of the primitive cell is somehow arbitrary [10]; for example, we can choose hexagonal area created by carbon-carbon bonds (see Figure 2b). The length of carbon-carbon bond is $d = 0.142\ nm$.

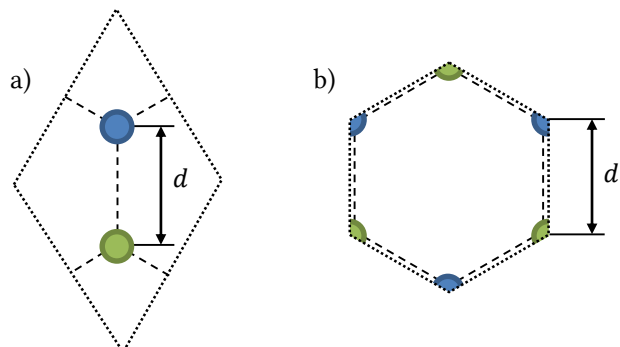

Figure 2. The primitive cells of graphene. Cell a) contains 2 atoms, each belonging to a different sublattice. Cell b) also contains ⅓ × 6 = 2 atoms.

The *Bravais lattice* vectors are shown on Figure 3. Their components are:

$$\vec{a}_1 = \begin{pmatrix} a \\ 0 \end{pmatrix}, \qquad \vec{a}_2 = \frac{1}{2}\begin{pmatrix} -a \\ \sqrt{3}a \end{pmatrix}$$

where $|\vec{a}_1| = |\vec{a}_2| = a = 0.246\ nm$. Having Bravais lattice vectors, one can derive the *reciprocal lattice* vectors:

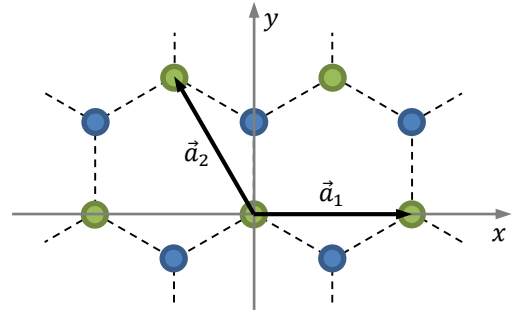

Figure 3. The Bravais lattice of graphene.

$$\vec{b}_1 = 2\pi \begin{pmatrix} 1 \\ 1/\sqrt{3} \end{pmatrix}\frac{1}{a}, \qquad \vec{b}_2 = 2\pi \begin{pmatrix} 0 \\ 2/\sqrt{3} \end{pmatrix}\frac{1}{a}$$

Using those, we can see that the *first Brillouin zone* is as shown on Figure 4. We have non-equivalent $K$-points: $K$ (marked in purple) and $K'$ (red), similarly to two sublattices of graphene. There is also $\Gamma$ point (blue) – centre of the reciprocal lattice cell, and $M$ point (orange) – in the middle of $K - K'$ bond.

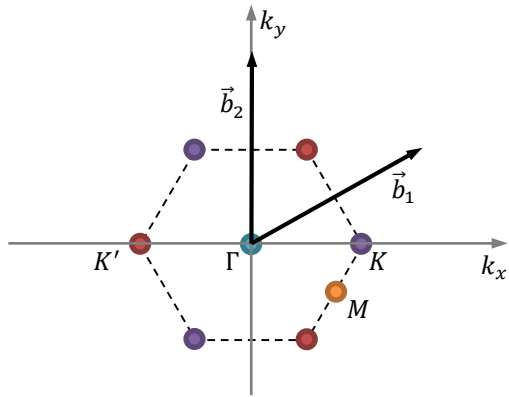

Figure 4. First Brillouin zone of graphene.

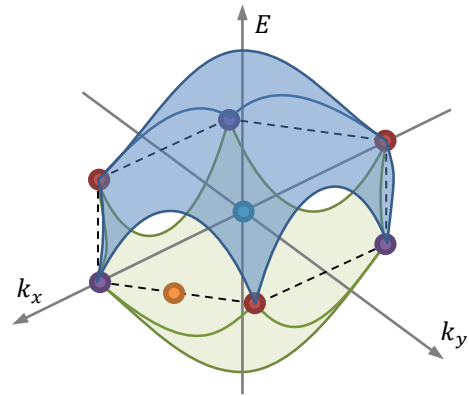

Figure 5. The band structure of graphene.

The *band structure* [11] of the graphene is shown on Figure 5. On the picture we can see two $\pi$ energy bands formed by p-orbitals, each one has the centre in the $K$-point. The bandgaps are open at the $M$-points, between the first and the second bands, while at the $K$-points we can see that they join. Energy bands at these points can be approximated by cones [12], as depicted on Figure 6. Area in blue represents area fully filled with electrons and we can see that Fermi level lies exactly in the $K$-point, that is the Dirac point. This is only true for not-doped graphene.

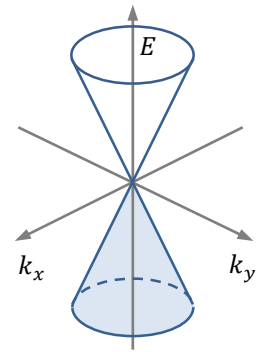

Figure 6. Energy near the corners $K, K'$ of the Brillouin zone.

What had to be also addressed are the various *applications* of graphene. Because of its unique electronic properties it may be used to create high quality transistors and electronic circuits [13]. Also, depending on how we cut nanoribbons from the graphene sheet, they can have a different electrical properties (for example can behave like semiconductors or like metals) [14]. They can be also used to create quantum dots [14] and thus may be useful in the future to create quantum computers. Because of its 2D structure, graphene is also thought to become a new material for manufacturing touchscreen monitors [3] or even flexible mobile phones.

### 1.1.2   Impurities on the graphene structure

If we spread some atoms or particles on the graphene lattice, then we deal with graphene impurities. Those impurities can change some properties of the graphene [15], so it is essential to know what impurities are present and how are they interacting. Impurities of one kind tend to occupy a special place on the lattice.

Figure 7 shows examples of how the impurities can choose their positions. Marked in blue, there is an impurity that sits in the middle of the hexagon. This is the most symmetric case. Red impurity sits on the carbon-carbon bond. The green case shows impurity that is placed on top of carbon atom – asymmetric position.

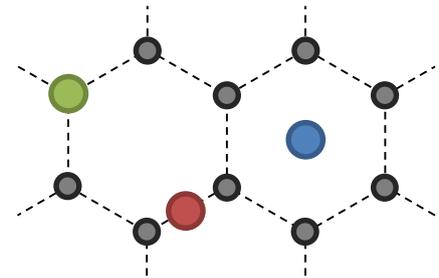

Figure 7. Graphene impurities occupying different positions.

Now, the properties of impurities are connected with on what sublattice the impurity sits [16]. For the atom sitting in the middle of the cell, we have three sublattices, marked on Figure 8a. For the case of impurity on the $C - C$ bond, we are also dealing with three sublattices, shown in Figure 8b. If the impurities sit on the top of $C$, we have two sublattices, corresponding to sublattices of the graphene (Figure 8c).

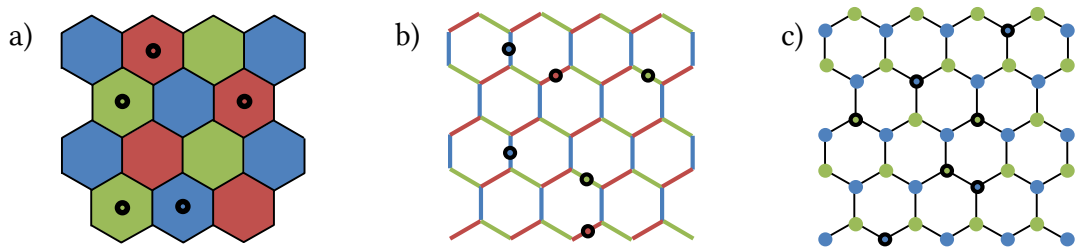

Figure 8. Different sublattices for different kinds of impurities in graphene.

If the impurities sit on the same sublattice, they will interact with each other differently, than if they would sit on the different sublattices [16]. It will be essential later, when we will be thinking of the algorithm to simulate our system.

## 1.2   Basics of Statistical Physics

In this section we will try to make the reader familiar with some (not all of course!) basics of statistical physics. They will be essential in the next chapter, when we will discuss the theoretical method used in our simulations.

Firstly, we have to introduce the concept of a statistical ensemble. Physical systems that we consider in thermodynamics often consists with a very large amount of particles. For example in one mole of a sample we will find $N_A = 6.02 \cdot 10^{23}$ atoms or molecules. This number is called Avogadro constant. Still, it is almost impossible to imagine the amount that it represents. The exact description of the system consisting with so many particles is impossible even with today's technology. (The total amount of global data is roughly estimated to be $10^{21}$ bytes [17]. If we would assume that one byte is sufficient to hold a data for a particle, this is still a hundred times too small to fully describe a mole of a substance.) This is why the statistical analysis must be introduced in description of these systems.

If we consider the system that is in finite temperature, it is obvious that the particles will have a kinetic energy. So this system will be constantly changing – particles will change their positions, sometimes they will hit each other and exchange momenta. **Microstate** of this system is just one particular "snapshot" taken in a particular moment of time. So, the state of the system would be defined as the sum of all microstates with corresponding weights. In a classical system, we have infinite number of microstates, because our variables (momenta, positions) are taking continuous values. But we will be considering only a case with a discrete values of energy, thus our number of microstates is finite.

**Statistical ensemble** is defined as all possible microstates of a system. We will be considering them all at once [18]. In addition to that, our system will have a constant number of particles, placed in a constant volume (or area) with a constant temperature. These constraints mean also that the system can exchange its energy with the surrounding. This particular case is called the *canonical ensemble* [19].

Each microstate $i$ with energy $E_i$ has a probability of occurrence given by *Boltzmann distribution* [18]:

$$p(E_i) = \frac{1}{Z} e^{-\beta E_i}, \qquad \beta = \frac{1}{k_B T}$$

where $k_B$ — Boltzmann constant, $T$ — temperature, $\beta$ — thermodynamic beta (introduced for simplicity), $Z$ — partition function (normalization factor).

**Partition function** (or **sum over states**) is very important quantity in statistical physics [18]. It is defined as:

$$Z \stackrel{\text{def}}{=} \sum_i e^{-\beta E_i} \tag{1.1}$$

The **statistical average** (mean value) of some property $A$, that depends on the energetic state of the system, is given by:

$$\langle A \rangle \stackrel{\text{def}}{=} \sum_i p_i A(E_i) = \frac{1}{Z} \sum_i A(E_i) e^{-\beta E_i} \tag{1.2}$$

This average is of macroscopic nature and can be measured by classical means, though it depends on microscopic states of the system. Our results generated by our algorithms will be mostly statistical averages.

What needs to be said at this point, is that sometimes we will use a name "Hamiltonian" to describe the energy. In fact, in classical mechanics *the Hamilton function*, or *the Hamiltonian* is a function describing the total energy of a system [20].

$$H = T + V$$

$T$ − kinetic energy, $V$ − potential energy. We will often use the Hamiltonian to denote the total energy of a microstate.

**Heat (thermal) capacity** is used to characterize how much heat is required to change a system's temperature by a specific amount. It is defined as [19]:

$$C \stackrel{\text{def}}{=} \frac{\partial U}{\partial T} = \frac{\partial \langle E \rangle}{\partial T} \tag{1.3}$$

where $U$ − internal energy. In case of the canonical ensemble, it is equal to mean energy. Also, we keep in this ensemble volume constant, so this would be *a heat capacity at constant volume* (denoted $C_V$ [10]), but for simplicity we will just call it "heat capacity" and denote with $C$ without subscript.

Another very useful quantity that we will measure is **magnetic susceptibility**. It characterizes how magnetization behaves in response to applied magnetic field. The *volume magnetic susceptibility* [21] $\chi$ is defined by the equation:

$$\vec{M} = \chi \vec{H}$$

where $\vec{M}$ − magnetization, $\vec{H}$ − magnetic field. Unfortunately in ferromagnetic material, this is not a linear relationship and we will have to use more general definition, sometimes called a *differential magnetic susceptibility* [22]:

$$\chi_{ij} \stackrel{\text{def}}{=} \frac{\partial M_i}{\partial H_j}$$

Here, magnetic susceptibility becomes a tensor (some kind of matrix). It has nine components, because $\vec{M}$ and $\vec{H}$ have three components each, $M_i, H_j$, where $i, j = 1,2,3$. Fortunately, in systems that are an object of consideration of this project, magnetization and magnetic field will have the same direction and will be denoted $M$ and $h$ (small letter to avoid the confusion with the Hamiltonian). Thus, we will be using only one component of susceptibility tensor, as the others are simply zeros.

$$\chi \stackrel{\text{def}}{=} \frac{\partial \langle M \rangle}{\partial h} \tag{1.4}$$

We used here $\langle M \rangle$ as this is the macroscopic quantity, that we really mean by previously used $M$.

## 1.3 Phase transition and the critical exponents

If the region of the substance has uniform properties, then we say that it is a certain *phase* of this system. For example, a different states of matter are also different phases – solids have different properties than liquids or gases. But if we have a sample of solid, then the physical properties are essentially uniform throughout the whole material.

The **phase transition** is the transformation between one phase of the system to another. For example, all processes like "melting", "freezing" or "condensation" are phase transitions. During the phase transition, the system changes some of its properties and sometimes it is happening discontinuously.

First noticeable **classification of the phase transitions** was introduced by Paul Ehrenfest [23]. It was based on the continuity of the chemical potential $\mu$. Namely, the phase transition is of $n$-th order, if the lowest derivative $\frac{\partial^n \mu}{\partial T^n}$ that has discontinuity during the transition is of $n$-th order. This classification is still commonly used, but it does not cover all transitions, and sometimes it is ambiguous.

Modern classification was proposed by Vitaly Ginzburg and Lev Landau and was based on Gibbs free energy $G$. There are two kinds of phase transition:

1) *Discontinuous phase transitions* – first derivative of $G$ is discontinuous and $G$ itself has a singularity in a form of spike.
2) *Continuous phase transitions* – $G$ is continuous and has continuous first derivatives. This means that there is no discontinuity in the specific heat $C$.

The *order parameter* $\mathcal{O}$ is the parameter that distinguishes two phases [23]. Usually, it is non-zero below the critical temperature $T_C$ (i.e. the temperature when the phase transition occurs) and vanished at and above $T_C$. Below the critical temperature, we have disordered phase and above there is ordered phase.

To describe the function $f(T)$ in the critical region (i.e. near the phase transition), we can introduce the dimensionless variable $t$, often called *reduced temperature* [23]:

$$t = \frac{T - T_C}{T_C} = \frac{T}{T_C} - 1$$

Our function becomes function of $t$: $f(t)$. We assume that the function is continuous near the phase transition (but not necessarily at the phase transition) and is positive. We may then define a limit $\lambda$ called the **critical exponent** [23]:

$$\lambda = \lim_{t \to 0} \frac{\ln f(t)}{\ln t}$$

We say, that this is the critical exponent of function $f(t)$ and write it in short form: $f(t) \sim t^\lambda$. This notation does not imply $f(t) = At^\lambda$. In general case there might be some **corrections-to-scaling** [23]:

$$f(t) = At^\lambda \left(1 + A_1 t^{\lambda_1} + A_2 t^{\lambda_2} + \cdots \right) \quad (1.5)$$

What is important, is that the critical exponents are believed to be universal – they only depend on the system dimension, spin dimension and range of interaction between them. It has not been proved yet, but the experimental data suggests this.

The most important critical exponents are shown in the table:

| Function | Disordered phase $t < 0$ | Ordered phase $t > 0$ |
|---|---|---|
| Specific heat $C$ | $C \sim (-t)^{-\alpha'}$ | $C \sim t^{-\alpha}$ |
| Order parameter $\mathcal{O}$ | $\mathcal{O} \sim (-t)^\beta$ | – |
| Susceptibility $\chi$ | $\chi \sim (-t)^{-\gamma'}$ | $\chi \sim t^{-\gamma}$ |
| Correlation length $\xi$ | $\xi \sim (-t)^{-\nu'}$ | $\xi \sim t^{-\nu}$ |

The susceptibility here is defined as $\partial \mathcal{O}/\partial h$, where $h$ is the source field. If the order parameter is magnetization and the source field is magnetic field, then of course $\chi$ is magnetic susceptibility. Correlation length $\xi$ describes the range of correlations in our system. We often introduce the *scaling condition*, that makes the exponents equivalent in both regions:

$$\alpha \equiv \alpha', \quad \gamma \equiv \gamma', \quad \nu \equiv \nu' \quad (1.6)$$

## 1.4 Ising model

Very important model of statistical system is the **Ising model**. In general, it consists of a constant number of sites, that are spread on a lattice or on a graph (continuous case) – see Figure 9. Each site bears a variable that can be in two states – this variable is called a spin.

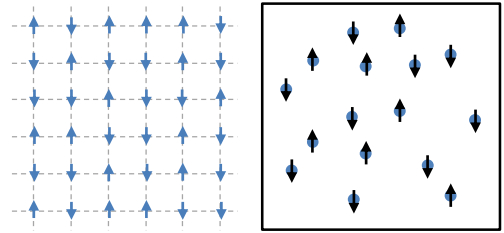

Figure 9. Examples of Ising model: on a square lattice and on a graph. Spins denoted by arrows.

*The Hamiltonian (energy)* of the Ising model is given by [19]:

$$H = -\sum_{i \neq j} J_{ij}\sigma_i\sigma_j - \sum_i h_i\sigma_i$$

where $i, j$ — denote the sites numbers, $J_{ij}$ — interaction constant, $\sigma_i, \sigma_j = \pm 1$ — spins, $h_i$ — value of magnetic field that acts on spin $i$. Spin variables change only the sign of the interaction constant. So, if we have $\sigma_i = \sigma_j$ spins are aligned in the same way (ferromagnetic) and we get $+1$, while if $\sigma_i = -\sigma_j$ spins are aligned in different directions (antiferromagnetic) and we get $-1$. This means that the sign of $J_{ij}$ is substantial:

1. If $J_{ij} > 0$, we have *ferromagnetic interaction* between spins $i$ and $j$. If the spins are aligned in the same way, the energy will decrease by the value of $J_{ij}$.
2. If $J_{ij} = 0$, we have *non-interacting spins*. Energy is not changing.
3. If $J_{ij} < 0$, we have *antiferromagnetic interaction* between spins. If the spins are aligned in opposite directions, the energy will decrease by the value of $J_{ij}$.

Very important case is the Ising model with *nearest neighbours interactions*. Spins here are only interacting with the nearest neighbours (most closely situated spins) and any other interaction is equal to zero. Also, it is common to assume that the interactions with the nearest neighbours are of the same value, equal to $J$. Furthermore, the magnetic field that acts on the spins is uniform and equal to $h$.

In this case our Hamiltonian is:

$$H = -J\sum_{\langle i,j \rangle} \sigma_i\sigma_j - h\sum_i \sigma_i \qquad (1.7)$$

where $\langle i, j \rangle$ — denotes the sites that are nearest neighbours.

It is also common to consider the case without any external magnetic field. Then the Hamiltonian (1.7) loses the second term:

$$H = -J\sum_{\langle i,j \rangle} \sigma_i\sigma_j \qquad (1.8)$$

But we have to remember, that when calculating the average of some variable directly connected to $h$, we must consider the full Hamiltonian first and then take the limit $h \to 0$.

The reason to consider the Ising model is to find if it shows any phase transition. *One dimensional case* was solved by Ernst Ising in 1925 [24]. It doesn't have a phase transition, so it will not be an object of our interest.

*Two-dimensional Ising model* was analytically solved by Lars Onsager in 1944 and proved to have the **phase transition** [19]. In fact 2D Ising model is the simplest physical model that shows the transition and is often given as an example of it. The critical temperature of this model is [19]:

$$T_C = \frac{2}{\ln(1 + \sqrt{2})} J \approx 2.269\,185\,314 \cdot J \tag{1.9}$$

What has to be also addressed is that the phase transition exists only in the thermodynamic limit [25], i.e. for systems with *infinite sizes*. For *finite size systems*, the phase transition will be smoothed out, but it still should be relatively easy to see. At this point we can already see that we should simulate not only one system size, but couple of them and investigate their behaviour in the critical region.

Another very important aspect is how to describe **the boundary conditions** of a system with a finite size. If we would consider a more physical case, with a system that has the edges "cut", then we can see that the outer spins would have less nearest neighbours then others. This will definitely change our results for a smaller system sizes. Because usually we are interested in the thermodynamic limit, it is common to introduce *periodic boundary conditions*, that are not changing our results so much. For a 1D case, this would mean, that the first and last site would be connected as a nearest neighbours, changing the topology to one of circle. For 2D case, we would have to consider a topology of an object that is a space product of two circles, i.e. a torus. Similarly, for higher dimensions, we would have a space product of $n$ circles, or so-called $n$-dimensional torus. In our research, we focus on 2D systems, so our topology is always that of torus.

# Chapter 2
# The details of the Monte Carlo methods

## 2.1 Classical Monte Carlo simulations and the Metropolis algorithm

Every computer simulation begins with a set of specific values – initial conditions. In our case, this would be the structure of our system (lattice/graph, number of sites) and value of temperature. Then we move on to a specific method of simulation.

**Monte Carlo methods** are simulation methods that rely on random, repeated updates of the system [22]. The method is based on the fact that by sampling microstates of the system with a certain probabilities, we can get a representative amount of data, that gives us approximate answer. Of course, the true answer lies in the limit of infinite running time.

Firstly, we have to discuss the process of update. There is a subgroup of methods called *random walk Monte Carlo*, or Markov chain Monte Carlo. The process used in these methods is often called **Metropolis update**, because Nicholas Metropolis along with others wrote in 1953 a paper proposing this algorithm [26]. The update consists with three simple steps:

1. Propose an update – change of the state $\nu$ to some new state $\nu'$.
2. Randomly accept or decline the update, with probability given by the balance equation.
3. Repeat this indefinitely.

This process creates a chain of states $\nu \to \nu' \to \nu'' \to \cdots$ that is called a *Markov chain* (see Figure 10) [21]. It is also called "random walk", because the system "walks" through all possible states. The big advantage of this method is that it is often very easy to construct a Markov chain that has a desired probability distribution.

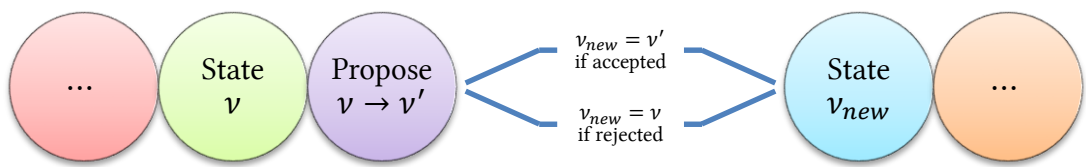

Figure 10. One "link" of the Markov chain.

We can also say, that basically the thermodynamic average that we discussed in **Chapter 1.2**, is replaced by the stochastic average. Thermodynamic sum over all configurations becomes the sum over randomly generated configurations. Of course, in Metropolis algorithm, these configurations are in fact getting a statistical weight, because of (mentioned before) balance equation. So, this stochastic average is giving us the same result in the limit of infinite calculation time.

But, as we now can see, there is one very important requirement, that must be satisfied. The two discussed sums (and averages) are equal to each other only if we will include all terms in the stochastic sum (obviously). It means that by using the Metropolis updates, we have to be able to generate from every initial state, any other state, in a limit of infinite simulation time. This is called **ergodicity** – if we run the simulation, it has to cover all configurations in the limit of infinite time [22].

We still didn't mentioned how to calculate the probabilities of accepting or rejecting the update. To do this, we need the **balance equation**. It is a very important formula that lies in the heart of all Monte Carlo methods [27]. It originates from the fact that the probability current of reaching the configuration $v$ must be equal to the probability current of going from $v$ to any other configuration:

$$W_v \sum_{v',u} p_u P_u^{\text{acc}}(v \to v') = \sum_{v',u} W_{v'} p_u P_u^{\text{acc}}(v' \to v) \tag{2.1}$$

where $W_v$ — weight of configuration $v$, $p_u$ — probability of suggesting update $u$ (and not any other update), $P_u^{\text{acc}}$ — probability that the update $u$ will be actually accepted (it is called the *acceptance probability*). The equation (2.1) reads, that the sum of probabilities to go from initial state to any other times the weight of this state must be equal to the sum of all probabilities to go from any other state back to the initial state times the weights of those other states. From this equation we can calculate how the acceptance ratio is given by the weight of configuration (usually given by the algorithm).

The easiest way to satisfy the balance equation (2.1) is to balance the updates in pairs – to make sure the probability current to perform update $u(v \to v')$ is exactly equal to the probability current to perform update $\bar{u}(v' \to v)$:

$$W_v p_u P_u^{\text{acc}}(v \to v') = W_{v'} p_{\bar{u}} P_{\bar{u}}^{\text{acc}}(v' \to v) \tag{2.2}$$

This equation is called the **detailed balance equation** [27]. In fact this simple equation represents a large amount of equations – every pair of updates $u, \bar{u}$ has its own detailed balance equation. This requirement obviously satisfies the general balance equation.

We can also define the quantity:

$$R = \frac{P_u^{\text{acc}}(v \to v')}{P_{\bar{u}}^{\text{acc}}(v' \to v)} = \frac{W_{v'}}{W_v} \frac{p_{\bar{u}}}{p_u} \tag{2.3}$$

that is called the *acceptance ratio* and will be often used in the description of our updates.

## 2.2 Problems and error estimation in MC algorithms

Of course, in the case of MC simulation that runs in the finite amount of time, we get the answer only approximately. Thus, we must think of how big the error should be, and how to calculate it appropriately.

Firstly, let us think if we should do the measurements of the variables, that we are interested in, after every update, or rather after some amount of updates. The Metropolis algorithm usually changes the system to the very similar state, with only small amount of values changed. Also, by looking at the Markov chain, we can see that the states are very dependent on the previous couple of states and they are correlated with them. So, if we want the measurement of some variable $A(v)$ to be accurately determined, we should measure its mean value $\langle A \rangle$ during the time, in which the system will "lose" the information about previous measurement. This time is called the **autocorrelation time** [27]. This time also indicates when the simulation covered a relevant region of configurations (i.e. those configurations that will be a representative sample of the whole configuration space).

So, after this autocorrelation time $t_0$, we get the result:

$$\langle A \rangle_1 \equiv \langle A \rangle_{t \in (0; t_0)} = \langle A \rangle_{\text{exact}} + \delta A_1$$

In the $n$-th "block" that will be measured between time $(n-1)t_0$ and $nt_0$ we get:

$$\langle A \rangle_n \equiv \langle A \rangle_{t \in ((n-1)t_0; nt_0)} = \langle A \rangle_{\text{exact}} + \underbrace{\delta A_n}_{\text{some random deviation}}$$

We can use now central limit theorem, because those averages $\langle A \rangle_n$ are not correlated, and say that our average over the entire simulation time is:

$$\langle A \rangle = \frac{\sum_{i=1}^{N} \langle A \rangle_i}{N}$$

and has the standard deviation:

$$\sigma = \frac{\sigma_{t_0}}{\sqrt{N}}$$

where $\sigma_{t_0}$ — standard deviation of the "block". We can now use the fact that the entire simulation time is $t = t_0 \cdot N$:

$$\sigma = \frac{\sigma_{t_0} \sqrt{t_0}}{\sqrt{t}} = \underbrace{\sigma_{t_0} \sqrt{t_0}}_{\text{constant}} \cdot \frac{1}{\sqrt{t}}$$

We can see, that by increasing the time of simulation, we can get more accurate results (obviously), but only as a factor of $1/\sqrt{t}$. We can calculate the error by using:

$$\sigma = \frac{1}{n}\sqrt{\sum_{i=1}^{N}(\langle A \rangle_i - \langle A \rangle)^2}$$

Unfortunately, in the whole analysis, we have an unknown parameter $t_0$. To deal with this problem we can introduce a special method – **binning method** [27]. The idea is to make bigger and bigger "blocks" by gluing together previously mentioned blocks and check how the errors behave.

We introduce the blocks numerators, that will be replacing the blocks averages:

$$\langle A \rangle_i = \frac{1}{Z_B}\sum_{\nu \in i\text{-th block}} A_\nu = \frac{R_i}{Z_B}$$

$\langle A \rangle_i$ – block averages, $R_i$ – block numerators, $Z_B$ – number of configurations in a block.

We form "superblocks" out from $j$ numbers of "blocks":

$$\langle B \rangle_i = \frac{1}{j}\sum_{k=(i-1)j+1}^{i \cdot j} \langle A \rangle_k = \frac{1}{j \cdot Z_B}\sum_{k=(i-1)j+1}^{i \cdot j} R_k$$

Number of these superblocks is obviously equal to $m = N/j$. The error in the blocking method is then given by:

$$\sigma^{(m)} = \frac{1}{m}\sqrt{\sum_{i=1}^{m}(\langle B \rangle_i - \langle A \rangle)^2}$$

We have to check how the errors behave when we make bigger and bigger blocks (increase $j$). Small blocks are correlates, so the errors should be smaller than their true value. But with big blocks we will not have correlated blocks and have a good value of our error.

Another phenomena that has to be addressed is the **thermalisation** problem. When we initialize the simulation, the system is not necessary near its thermal equilibrium. So, during the time it is thermalizing, the measured values should not be considered, because their values are mainly influenced by the initial condition and the algorithm. Therefore, it is preferred to start collecting the data after the thermalisation time. We say then, that the system is "well thermalized".

The problem, that we will also encounter trying to simulate the graphene impurities, is the **sign problem** [27]. Let us consider a system that may have weights $W_\nu$ both positive and negative. Then the average of some variable $\langle A \rangle$ is:

$$\langle A \rangle = \frac{\sum_\nu A(\nu) W_\nu}{\sum_\nu W_\nu} = \frac{\sum_\nu A(\nu) Sign_\nu |W_\nu|}{\sum_\nu Sign_\nu |W_\nu|}$$

If we now change this sum with positive weights $|W_\nu|$ to stochastic sum we get:

$$\langle A \rangle = \frac{\sum_\nu^{sto} A(\nu) Sign_\nu}{\sum_\nu^{sto} Sign_\nu} = \frac{\langle A \cdot Sign \rangle}{\langle Sign \rangle}$$

If the sign is changing, we may get $\langle Sign \rangle \to 0$. This would give us very bad results, as it would enlarge the errors largely. Fortunately, we can often change the representation of the system's Hamiltonian, so that it would have no sign problem.

## 2.3 Worm algorithm

### 2.3.1 Using worm algorithm on Ising model

The worm algorithm (WA) was firstly introduced by Nikolay Prokof'ev and Boris Svistunov in 2001 in high-temperature expansion and was shown to have an amazing performance in this region [28].

We introduce the worm algorithm (WA) in an example of Ising model [27]. Firstly, we write down the Hamiltonian for this model, using equation (1.8):

$$H = -J \sum_{\langle i,j \rangle} \sigma_i \sigma_j = -J \sum_b \sigma_i \sigma_{i+\mu} \tag{2.4}$$

where $J$ — coupling constant and $\langle i,j \rangle$ — means that the sites $i$ and $j$ are the nearest neighbours. This notation is changed to another, where $b = (i, \mu)$ — bonds between the site $i$ and nearest neighbours of this site, which are denoted here as $(i + \mu)$.

We use this to calculate the partition function using the definition (1.1) and inserting equation (2.4):

$$Z = \sum_{\{\sigma_i\}} e^{-\beta H} = \sum_{\{\sigma_i\}} e^{K \sum_b \sigma_i \sigma_{i+\mu}} = \sum_{\{\sigma_i\}} \prod_b e^{K \sigma_i \sigma_{i+\mu}} \tag{2.5}$$

where $\{\sigma_i\}$ — denotes all possible configurations, $K = \beta J$.

Now, we use the fact, that we can Taylor-expand the exponential:

$$e^{K \sigma_i \sigma_{i+\mu}} = \sum_{n_b=0}^{\infty} \frac{(K \sigma_i \sigma_{i+\mu})^{n_b}}{n_b!} \tag{2.6}$$

Here, $n_b$ is just some variable in Taylor-expansion process, but it also has a deeper meaning. Every bond $b$ has this number. When representing this algorithm in a picture form, we will use this number as indicator, how many times the path crosses a particular bond. Thus, the process of "drawing" will simply mean that we have to increase $n_b$ and similarly the process of "erasing" is decreasing.

This expansion is called *high-temperature expansion for Ising model*, because the performance of the algorithm is very high, but only for high temperatures, much like in the original paper [28].

Formula (2.5) for partition function becomes:

$$Z = \sum_{\{\sigma_i\}} \prod_b \sum_{n_b=0}^{\infty} \frac{K^{n_b}(\sigma_i \sigma_{i+\mu})^{n_b}}{n_b!}$$

If we expand this, we can see that:

$$Z = \sum_{\{\sigma_i\}} \sum_{\{n_b\}} \frac{K^{n_1}}{n_1!} \frac{K^{n_2}}{n_2!} \frac{K^{n_3}}{n_3!} \cdots (\sigma_{i_1}\sigma_{i_1+\mu_1})^{n_1} (\sigma_{i_2}\sigma_{i_2+\mu_2})^{n_2} \cdots$$

$$= \sum_{\{n_b\}} \left( \prod_b \frac{K^{n_b}}{n_b!} \right) \left( \sum_{\sigma_1=\pm1} \sigma_1^{\sum_\mu n_{(1,\mu)}} \sum_{\sigma_2=\pm1} \sigma_2^{\sum_\mu n_{(2,\mu)}} \cdots \right)$$

$$= \sum_{\{n_b\}} \left( \prod_b \frac{K^{n_b}}{n_b!} \right) \left( \prod_{i=1}^{N} \sum_{\sigma_i=\pm1} \sigma_i^{p_i} \right)$$

In the last equality we defined $p_i = \sum_\mu n_{(i,\mu)}$. Now, one can see that depending if $p_i$ is even or odd, the last sum vanishes or becomes number 2:

$$\sum_{\sigma_i=\pm1} \sigma_i^{p_i} = \begin{cases} 1-1 & \text{if } p_i \text{ is odd} \\ 1+1 & \text{if } p_i \text{ is even} \end{cases} = \begin{cases} 0 & \text{if } p_i \text{ is odd} \\ 2 & \text{if } p_i \text{ is even} \end{cases}$$

$$\Rightarrow \prod_{i=1}^{N} \sum_{\sigma_i=\pm1} \sigma_i^{p_i} = \begin{cases} 0 & \text{if } p_i \text{ is odd} \\ 2^N & \text{if } p_i \text{ is even} \end{cases}$$

Thus, only those terms, where $p_i$ is even, are contributing to the overall $Z$. $p_i$ is the number of bonds between site $i$ and the nearest neighbours. The constraint "$p_i$ is even" means that the path created by the bonds is always closed.

$$Z = \sum_{\{n_b\}_{CP}} \left( 2^N \prod_b \frac{K^{n_b}}{n_b!} \right) \equiv \sum_{\{n_b\}_{CP}} W(\{n_b\}) \qquad (2.7)$$

where $\{n_b\}_{CP}$ means the all possible combinations of bonds, that are graphically represented by a closed paths (CP). We denoted the product times $2^N$ as $W(\{n_b\})$ as it is simply some weight that corresponds to a given configuration $\{n_b\}$.

Obtained equation (2.7) is very important, as it shows how to perform update in the worm algorithm. But before writing the algorithm itself, we should calculate the acceptance ratio. To do it, we use the detailed balance equation (2.2) in slightly condensed form:

$$P^{\text{acc}}_{v \to v'} p_u W_v = P^{\text{acc}}_{v' \to v} p_{\bar{u}} W_{v'}$$

Let's say that we want to change one of $n_b$, we will denote it $n_c$, to be $n_c \to n_c + 1$. This operation will be referred as "drawing", because it "draws" additional bond. Then:

$$P^{\text{acc}}_{n_c \to n_c+1} \frac{1}{2}\frac{1}{d} W(\{n_b\}_{\text{initial}}) = P^{\text{acc}}_{n_c+1 \to n_c} \frac{1}{2}\frac{1}{d} W(\{n_b\}_{\text{final}})$$

Here, $p_u = \frac{1}{2}\frac{1}{d}$ was used here, because one firstly is choosing if to draw or erase, and then selecting one of $d$ nearest neighbours. Expanding $W$ using (2.7) gives:

$$P^{\text{acc}}_{n_c \to n_c+1} 2^N \frac{K^{n_1}}{n_1!}\frac{K^{n_2}}{n_2!}\frac{K^{n_3}}{n_3!}\cdots\frac{K^{n_c}}{n_c!}\cdots = P^{\text{acc}}_{n_c+1 \to n_c} 2^N \frac{K^{n_1}}{n_1!}\frac{K^{n_2}}{n_2!}\frac{K^{n_3}}{n_3!}\cdots\frac{K^{n_c+1}}{(n_c+1)!}\cdots$$

$$P^{\text{acc}}_{n_c \to n_c+1} \frac{K^{n_1}}{n_1!}\frac{K^{n_2}}{n_2!}\frac{K^{n_3}}{n_3!}\cdots\frac{K^{n_c}}{n_c!}\cdots = P^{\text{acc}}_{n_c+1 \to n_c} \frac{K^{n_1}}{n_1!}\frac{K^{n_2}}{n_2!}\frac{K^{n_3}}{n_3!}\cdots\frac{K^{n_c} \cdot K}{n_c! \cdot (n_c+1)}\cdots$$

$$P^{\text{acc}}_{n_c \to n_c+1} \frac{K^{n_1}}{n_1!}\frac{K^{n_2}}{n_2!}\frac{K^{n_3}}{n_3!}\cdots\frac{K^{n_c}}{n_c!}\cdots = P^{\text{acc}}_{n_c+1 \to n_c} \frac{K}{n_c+1}\cdot\frac{K^{n_1}}{n_1!}\frac{K^{n_2}}{n_2!}\frac{K^{n_3}}{n_3!}\cdots\frac{K^{n_c}}{n_c!}\cdots$$

That gives rise to the acceptance ratio (2.3) given by:

$$R \equiv \frac{P^{\text{acc}}_{n_c \to n_c+1}}{P^{\text{acc}}_{n_c+1 \to n_c}} = \frac{K}{n_c+1}$$

Now, the "erasing" operation will be very similar ($n_c \to n_c - 1$):

$$P^{\text{acc}}_{n_c \to n_c-1} \frac{K^{n_1}}{n_1!}\frac{K^{n_2}}{n_2!}\frac{K^{n_3}}{n_3!}\cdots\frac{K^{n_c}}{n_c!}\cdots = P^{\text{acc}}_{n_c-1 \to n_c} \frac{K^{n_1}}{n_1!}\frac{K^{n_2}}{n_2!}\frac{K^{n_3}}{n_3!}\cdots\frac{K^{n_c-1}}{(n_c-1)!}\cdots$$

$$P^{\text{acc}}_{n_c \to n_c-1} \frac{K^{n_1}}{n_1!}\frac{K^{n_2}}{n_2!}\frac{K^{n_3}}{n_3!}\cdots\frac{K^{n_c}}{n_c!}\cdots = P^{\text{acc}}_{n_c-1 \to n_c} \frac{n_c}{K}\cdot\frac{K^{n_1}}{n_1!}\frac{K^{n_2}}{n_2!}\frac{K^{n_3}}{n_3!}\cdots\frac{K^{n_c}}{n_c!}\cdots$$

So:

$$R \equiv \frac{P^{\text{acc}}_{n_c \to n_c-1}}{P^{\text{acc}}_{n_c-1 \to n_c}} = \frac{n_c}{K}$$

If one keep track of every bond number $n_b$, then calculating the acceptance ratio for Metropolis update is fairly simple:

$$R = \begin{cases} K/(n_b+1) & \text{if drawing} \\ n_b/K & \text{if erasing} \end{cases} \qquad (2.8)$$

### 2.3.2 WA in picture representation and the algorithm

It is noticeable that our description of worm algorithm uses many art terms such as "drawing" or "erasing". This is used for better understanding of this algorithm and so that we can actually imagine how it works.

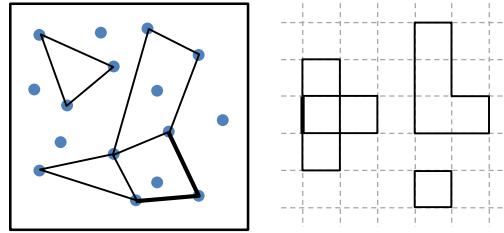

Figure 11. CP configurations for a graph and a square lattice.

Our system can be considered as a lattice or a graph, with some number of sites. Sometimes there is also a continuous case, where we have an infinite number of sites forming a figure, but this is not object of our investigation. Now, closed path configurations are represented as shown on Figure 11. Thinner lines represent single bonds ($n_b = 1$), thicker lines indicate double bonds ($n_b = 2$) and of course no bond corresponds to $n_b = 0$. We remember that CP was concluded from the fact, that every site had to have an even number of bonds connected to it. We can see that now on the picture.

The "drawing" process was defined to be $n_b \to n_b + 1$. We can see that now, that it simply means that we draw additional bond on our diagram. Similarly, the "erasing" process is just erasing one of the bonds.

We are now prepared to write the worm algorithm.

1) We have to start in some particular site. Let us choose this site randomly to be $j_1 = j_2$. It is conventional to call the starting point "ira" and ending point of our drawn path "masha". Of course in the beginning ("ira" = $j_1$) = ("masha" = $j_2$).
2) Select randomly the proposed new site $j_3$. If we use square lattice Ising model with nearest neighbours interactions, there are only four possibilities. If we have a graph where every site can interact with every other, then we have $N_{sites}$ possibilities.
3) Choose randomly, if we want to draw or to erase the bond between bonds $j_2$ and $j_3$.
4) Perform the "shift" (i.e. move "masha" to $j_3$ and draw/erase, as proposed) only if random number from region $(0,1)$ is less than the acceptance ratio $R$ (we have to calculate it using (2.8) ). So, if accepted, our $j_3$ becomes new "masha" = $j_2$.
5) Go back to 2) if "ira" ≠ "masha". If they are equal, then choose new site, as described in 1).

Every time when we *retrieve the CP configuration* (i.e. when we have "ira" = "masha"), then using equation (2.7), we can see that we should add $+1$ to our measurement of $Z$. Also, we should perform all the measurements of variables, that means of we are interested in.

### 2.3.3 Limiting the configuration space

As we can see, we are dealing here with the infinite configuration space. I.e. the bond number can take any positive integer value. This is due to Taylor expansion of the exponent in the equation (2.6). There is another way to expand this exponent, simply using the fact that $\sigma_i \sigma_j = \pm 1$, so it has only two possible values [27]:

$$e^{K\sigma_i \sigma_j} = \cosh K + \sigma_i \sigma_j \sinh K = \cosh K \cdot \sum_{n_b=0}^{1} (\sigma_i \sigma_j)^{n_b} \tanh^{n_b} K$$

We have just limited the possible values of bond number from infinity to two. This case is extremely important if we want to do the simulation of low temperatures. The algorithm gets a boost in performance, because when $T \to 0 \Rightarrow \tanh K \to 1$.

The algorithm is exactly the same as before, except we have a different acceptance ratio:

$$R = \begin{cases} \tanh K & \text{if drawing} \\ \tanh^{-1} K & \text{if erasing} \end{cases} \tag{2.9}$$

### 2.3.4 Measurements in worm algorithm

We now have to calculate the variables that we want to keep track of. As was mentioned before, we are interested in phase transition. The variables that will show some unusual behaviour in the critical region are in particular magnetic susceptibility and heat capacity.

Firstly, let us calculate $\chi$. Using the definition of **magnetic susceptibility** (1.4):

$$\chi \stackrel{\text{def}}{=} \frac{\partial \langle M \rangle}{\partial h}$$

So, firstly one has to obtain the equation for mean magnetization. This is possible by using the full equation for energy in Ising model (1.7):

$$H = -J \sum_{\langle i,j \rangle} \sigma_i \sigma_j - h \underbrace{\sum_i \sigma_i}_{\text{magnetization}} = -J \sum_{\langle i,j \rangle} \sigma_i \sigma_j - h\,M$$

If we calculate the partial derivative of $Z$ with respect to the field $h$:

$$\frac{\partial Z}{\partial h} = \frac{\partial}{\partial h} \left( \sum_{\{\sigma_i\}} e^{-\beta E} \right) = \sum_{\{\sigma_i\}} (-\beta)(-M) e^{-\beta E} = \beta Z \cdot \frac{1}{Z} \sum_{\{\sigma_i\}} M e^{-\beta E} = \beta Z \langle M \rangle$$

$$\langle M \rangle = \frac{1}{\beta Z} \frac{\partial Z}{\partial h} \tag{2.10}$$

Using this, we obtain a better form for $\chi$:

$$\chi = \frac{\partial \langle M \rangle}{\partial h} = \frac{1}{\beta}\frac{\partial}{\partial h}\left(\frac{1}{Z}\frac{\partial Z}{\partial h}\right) = \frac{1}{\beta}\left(-\frac{1}{Z^2}\frac{\partial Z}{\partial h}\frac{\partial Z}{\partial h} + \frac{1}{Z}\frac{\partial^2 Z}{\partial h^2}\right) = \frac{1}{\beta}\left(\frac{1}{Z}\frac{\partial^2 Z}{\partial h^2} - \left(\frac{1}{Z}\frac{\partial Z}{\partial h}\right)^2\right)$$

From the equation (2.10) we obtain the second term. The first term must be done explicitly:

$$\frac{1}{Z}\frac{\partial^2 Z}{\partial h^2} = \frac{1}{Z}\frac{\partial}{\partial h}\left(\frac{\partial Z}{\partial h}\right) = \frac{1}{Z}\frac{\partial}{\partial h}\left(\beta \sum_{\{\sigma_i\}} M e^{-\beta E}\right) = \frac{1}{Z}\beta^2 \sum_{\{\sigma_i\}} M^2 e^{-\beta E} = \beta^2 \langle M^2 \rangle$$

Finally, we get:

$$\chi = \frac{1}{\beta}(\beta^2 \langle M^2 \rangle - \beta^2 \langle M \rangle^2) = \beta(\langle M^2 \rangle - \langle M \rangle^2)$$

Now, if we use the fact that: $M = \sum_i \sigma_i$ we get $M^2 = \sum_{i,j} \sigma_i \sigma_j$. In our model, there is no external magnetic field. So $\langle M \rangle = 0$. Therefore:

$$\chi = \beta \langle M^2 \rangle = \beta \frac{1}{Z} \sum_{\{\sigma_k\}} \sum_{i,j} \sigma_i \sigma_j \, e^{K \sum_b \sigma_k \sigma_{k+\mu}} = \beta \sum_{i,j} \underbrace{\frac{1}{Z} \sum_{\{\sigma_k\}} \sigma_i \sigma_j e^{K \sum_b \sigma_k \sigma_{k+\mu}}}_{\text{spin-spin correlation function}}$$

$$\chi = \beta \sum_{i,j} g(i-j) \qquad (2.11)$$

We used here the quantity named **spin-spin correlation function**. It tells us how one spin (its direction) influences all others. Its formal definition is [23]:

$$g(i-j) \stackrel{\text{def}}{=} \langle \vec{\sigma}_i \cdot \vec{\sigma}_j \rangle \stackrel{\text{our case}}{=} \langle \sigma_i \sigma_j \rangle \qquad (2.12)$$

We use this equation to calculate it in WA representation:

$$g(j_1 - j_2) = \frac{1}{Z} \sum_{\{\sigma_i\}} \sigma_{j_1} \sigma_{j_2} e^{K \sum_b \sigma_i \sigma_{i+\mu}} = \frac{1}{Z} \sum_{\{\sigma_i\}} \sigma_{j_1} \sigma_{j_2} \prod_b \sum_{n_b=0}^{\infty} \frac{K^{n_b}(\sigma_i \sigma_{i+\mu})^{n_b}}{n_b!}$$

$$= \frac{1}{Z} \sum_{\{n_b\}} \left(\prod_b \frac{K^{n_b}}{n_b!}\right) \sum_{\sigma_1=\pm 1} \sigma_1^{\sum_\mu n_{(1,\mu)}} \cdots \sum_{\sigma_{j_1}=\pm 1} \sigma_{j_1}^{\sum_\mu n_{(j_1,\mu)}+1} \sum_{\sigma_1=\pm 1} \sigma_{j_2}^{\sum_\mu n_{(j_2,\mu)}+1} \cdots$$

$$= \frac{1}{Z} \sum_{\{n_b\}} \left(\prod_b \frac{K^{n_b}}{n_b!}\right) \prod_{\substack{i=1 \\ i \neq j_1 \\ i \neq j_2}}^{N} \left(\sum_{\sigma_i=\pm 1} \sigma_i^{p_i}\right) \cdot \sum_{\sigma_{j_1}=\pm 1} \sigma_{j_1}^{p_{j_1}+1} \cdot \sum_{\sigma_{j_2}=\pm 1} \sigma_{j_2}^{p_{j_2}+1}$$

$$\sum_{\sigma_i=\pm 1} \sigma_i^{p_i} = \begin{cases} 0 & \text{if } p_i \text{ is odd} \\ 2 & \text{if } p_i \text{ is even} \end{cases}$$

$$\sum_{\sigma_{j_1}=\pm 1} \sigma_{j_1}^{p_{j_1}+1} = \begin{cases} 2 & \text{if } p_{j_1} \text{ is odd} \\ 0 & \text{if } p_{j_1} \text{ is even} \end{cases}, \qquad \sum_{\sigma_{j_2}=\pm 1} \sigma_{j_2}^{p_{j_2}+1} = \begin{cases} 2 & \text{if } p_{j_2} \text{ is odd} \\ 0 & \text{if } p_{j_2} \text{ is even} \end{cases}$$

So, if we assume that we have a closed path, but with two unconnected edges, $j_1, j_2$, then we obtain:

$$g(j_1 - j_2) = \frac{1}{Z} \sum_{\{n_b\}_{CP_g}} W(\{n_b\}) \equiv \frac{G(j_1 - j_2)}{Z} \tag{2.13}$$

where $CP_g$ is a symbol for closed path with two edges unconnected. So, if we want to measure the magnetic susceptibility using (2.11), we have keep track of $\sum_{i,j} G(i-j) \equiv G$ (as we already know how to keep track of $Z$). This time, every configuration is contributing to $G$. So, similarly to $Z$, we add $+1$ to our variable $G$, but for *every configuration* (as opposed to $Z$, which was measured only when we had CP configuration).

Calculating the **heat capacity**. Using the definition (1.3):

$$C \stackrel{\text{def}}{=} \frac{\partial \langle E \rangle}{\partial T}$$

We remember that:

$$\langle E \rangle = -\frac{\partial}{\partial \beta} \ln Z = -\frac{1}{Z} \frac{\partial Z}{\partial \beta} \tag{2.14}$$

Firstly, we change the differentiation variable:

$$T = \frac{1}{\beta} \Rightarrow \frac{\partial T}{\partial \beta} = -\frac{1}{\beta^2} \Rightarrow \frac{\partial}{\partial T} = -\beta^2 \frac{\partial}{\partial \beta}$$

Therefore:

$$C = -\beta^2 \frac{\partial \langle E \rangle}{\partial \beta} = \beta^2 \frac{\partial}{\partial \beta} \left( \frac{1}{Z} \frac{\partial Z}{\partial \beta} \right) = \beta^2 \left( -\frac{1}{Z^2} \frac{\partial Z}{\partial \beta} \frac{\partial Z}{\partial \beta} + \frac{1}{Z} \frac{\partial^2 Z}{\partial \beta^2} \right)$$
$$= \beta^2 \left( \frac{1}{Z} \frac{\partial^2 Z}{\partial \beta^2} - \left( \frac{1}{Z} \frac{\partial Z}{\partial \beta} \right)^2 \right)$$

Equation (2.14) gives us the second term. The first term is:

$$\frac{1}{Z} \frac{\partial}{\partial \beta} \left( \frac{\partial Z}{\partial \beta} \right) = \frac{1}{Z} \frac{\partial}{\partial \beta} \left( -\sum_{\{\sigma_i\}} E\, e^{-\beta E} \right) = \frac{1}{Z} \sum_{\{\sigma_i\}} E^2 e^{-\beta E} = \langle E^2 \rangle \tag{2.15}$$

Thus we obtain:

$$C = \beta^2 (\langle E^2 \rangle - \langle E \rangle^2) \tag{2.16}$$

that is very similar to the result for $\chi$. To calculate $\langle E \rangle$ and $\langle E^2 \rangle$ we can use equations (2.14) and (2.15), and the representation of partition function for worm algorithm (2.7):

$$Z = \sum_{\{n_b\}_{CP}} W(\{n_b\}) = 2^N \sum_{\{n_b\}_{CP}} \prod_b \frac{K^{n_b}}{n_b!} = 2^N \sum_{\{n_b\}_{CP}} \frac{K^{n_1+n_2+n_3+\cdots}}{n_1!\, n_2!\, n_3!\, \ldots}, \qquad K = \beta J$$

So:

$$\langle E \rangle = -\frac{1}{Z}\frac{\partial Z}{\partial \beta} = -\frac{1}{Z} 2^N \sum_{\{n_b\}_{CP}} \frac{(n_1+n_2+n_3+\cdots) K^{n_1+n_2+n_3+\cdots-1}}{n_1!\, n_2!\, n_3!\, \ldots} \cdot J$$

$$= -\frac{1}{Z} 2^N \frac{J}{K} \sum_{\{n_b\}_{CP}} \frac{(\sum_b n_b) K^{n_1+n_2+n_3+\cdots}}{n_1!\, n_2!\, n_3!\, \ldots} = -\frac{1}{\beta}\frac{1}{Z}\sum_{\{n_b\}_{CP}} N_b W(\{n_b\})$$

$$\langle E \rangle = -\frac{1}{\beta}\langle N_b \rangle \qquad (2.17)$$

Where we denoted $N_b = \sum_b n_b$ that is the sum of all bond numbers. This quantity is easy to keep track of, because we simply add 1 to it, when we draw and subtract 1, when we erase. We can also see that in the meantime we got the equation (2.17) for the **mean energy** measurement. Now, we can calculate $\langle E^2 \rangle$:

$$\langle E^2 \rangle = \frac{1}{Z}\frac{\partial^2 Z}{\partial \beta^2} = \frac{1}{Z}\frac{\partial}{\partial \beta}\left( \frac{1}{\beta}\sum_{\{n_b\}_{CP}} N_b W(\{n_b\}) \right) = \frac{1}{Z}\sum_{\{n_b\}_{CP}} N_b \frac{\partial}{\partial \beta}\left( \frac{1}{\beta} W(\{n_b\}) \right)$$

$$= \frac{1}{Z}\sum_{\{n_b\}_{CP}} N_b \left( -\frac{1}{\beta^2} W(\{n_b\}) + \frac{1}{\beta}\cdot\frac{1}{\beta} N_b W(\{n_b\}) \right)$$

$$= \frac{1}{\beta^2}\frac{1}{Z}\sum_{\{n_b\}_{CP}} (N_b^2 - N_b) W(\{n_b\}) = \frac{1}{\beta^2}\langle N_b^2 - N_b \rangle$$

So, *every time we retrieve the close path configuration*, we have to measure $N_b$ and $(N_b^2 - N_b)$. At the end of our simulation we have to evaluate $C$ using (2.16):

$$C = \beta^2(\langle E^2 \rangle - \langle E \rangle^2) = \beta^2 \left( \frac{1}{\beta^2}\langle N_b^2 - N_b \rangle - \frac{1}{\beta^2}\langle N_b \rangle^2 \right)$$

$$C = \langle N_b^2 - N_b \rangle - \langle N_b \rangle^2 \qquad (2.18)$$

# Chapter 3
# Simulation of the Ising model

## 3.1 Preparing the computer simulation and results

The simulation that was set up, was using previously introduced in **Chapter 2.3.1** worm algorithm for the Ising model described by the Hamiltonian (1.8). Our object of interest is the case of positive interaction constant $J > 0$, as the worm algorithm works only for the ferromagnetic systems.

As said before, to accurately describe the behaviour of the system with infinite size, we have to investigate a finite size systems. In our simulation we used three different systems:

| System size $L$ | Sweeps number | Thermalisation |
|---|---|---|
| 10 | 1 000 000 | 50 000 |
| 50 | 10 000 | 1 000 |
| 100 | 10 000 | 1 000 |

$L$ describes the size of the system: $L \times L$ is the number of spins that are distributed on the square lattice. Sweeps number is the total number of updates of the system, while "thermalisation" column describes the number of sweeps after which we consider the system to be "well thermalized" and start to collect measurements.

Now, we have to set up, which range of temperatures our simulation should check.

| $L$ | $T/J$ | | | |
|---|---|---|---|---|
| | Range | Sampling | Critical region (CR) | Sampling in CR |
| 10 | [0.1; 5.0] | 0.10 | [2.0; 3.0] | 0.050 |
| 50 | [1.5; 3.0] | 0.05 | [2.1; 2.4] | 0.025 |
| 100 | N/A | N/A | [2.1; 2.4] | 0.025 |

Because the $L = 10$ system was simulated before others as a check of algorithm efficiency, it has a larger range. The sampling in critical region is bigger, but for the purpose of finding the critical temperature, it was improved later. For the system with $L = 100$ only the simulation of the critical region was performed.

Our results can be shown in form of plots – see Figure 12 and Figure 13.

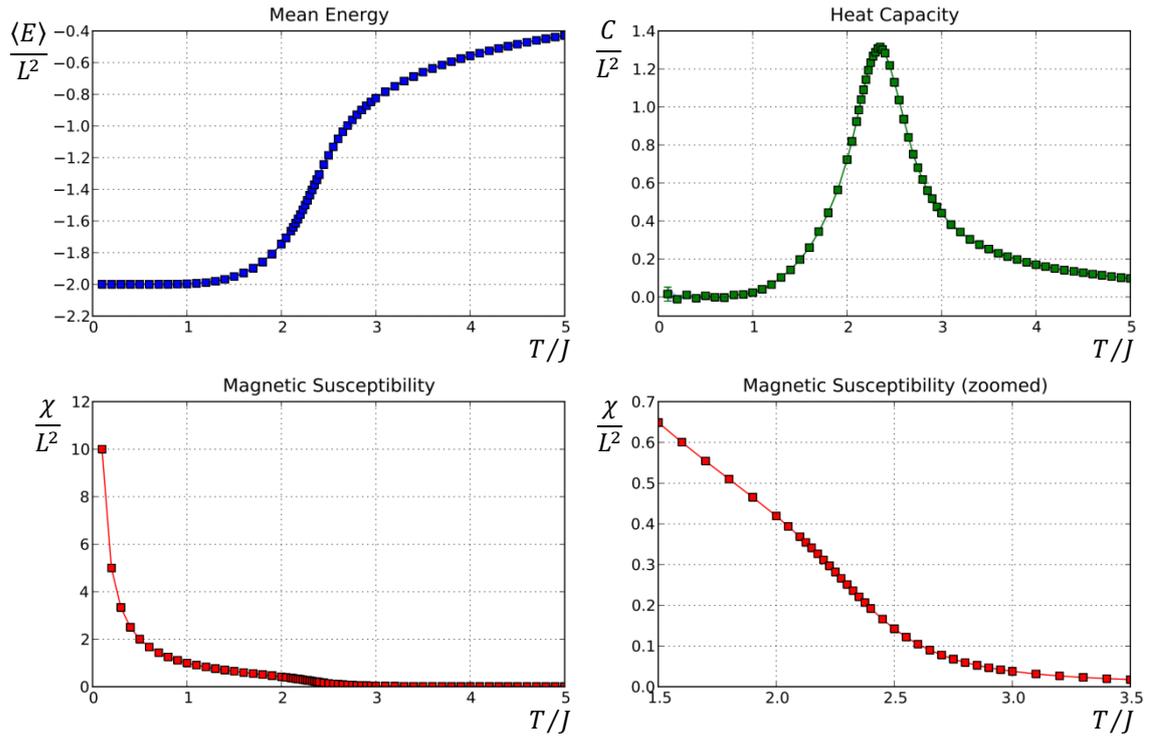

Figure 12. Mean energy, heat capacity and magnetic susceptibility for system with size $L = 10$. All the results are normalized (i.e. they represent values per one site).

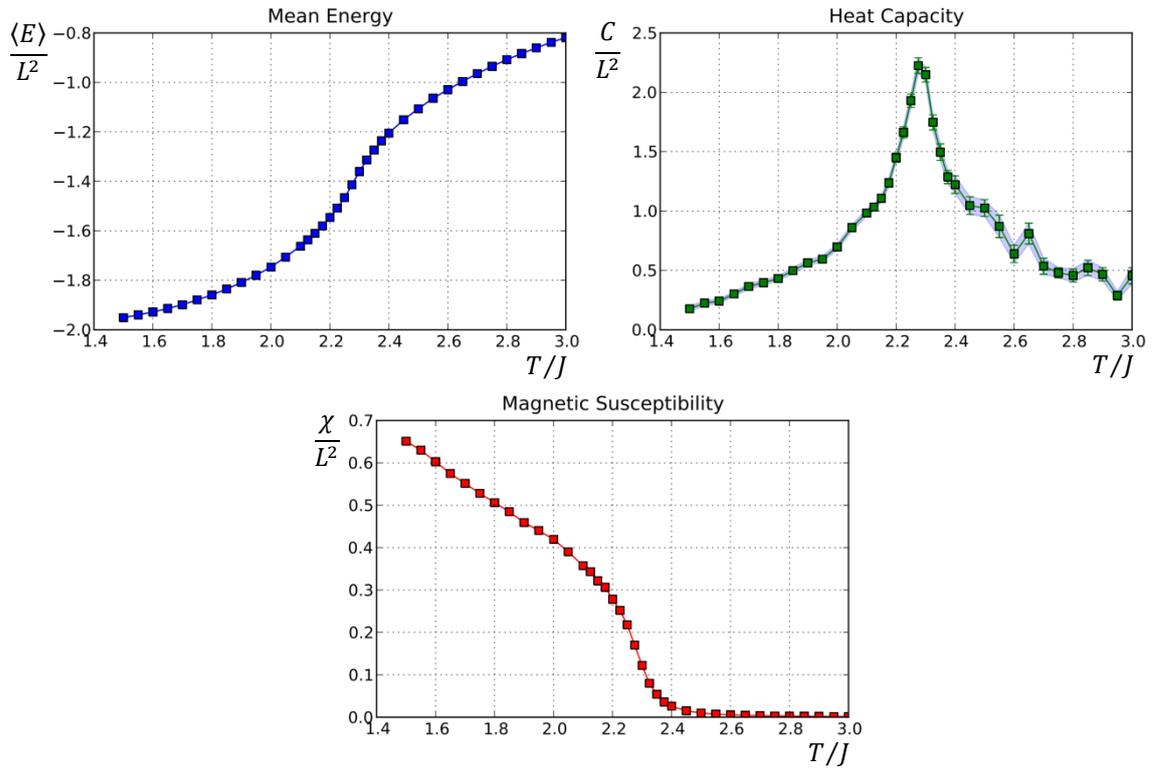

Figure 13. Mean energy, heat capacity and magnetic susceptibility for system with size $L = 50$.

The results for the system size $L = 100$ are not shown on the plot, because the region covered is substantially smaller than for the rest. Those results were only used for finding the critical temperature.

## 3.2 Analysis of the results

As we can see from the plots, the critical temperature is somewhere in the range $T/J \in (2.2; 2.4)$. Mean energy has a change of sign of the second derivative as well as magnetic susceptibility, while the heat capacity shows the peak in the critical region.

To better estimate the value of critical temperature, we can use the following idea. Using the critical exponents (**Chapter 1.3**), we write that the order parameter (in our case it is magnetization per site, $m$) below the critical temperature is proportional to:

$$m \sim (-t)^\beta \quad \Rightarrow \quad \langle m^2 \rangle \sim (-t)^{2\beta}$$

The variables that can describe our $\langle m^2 \rangle$ can only be temperature (that is included in $t$), the size of the system $L$ and the correlation length $\xi$. Now, we postulate, that:

$$\langle m^2 \rangle = (-t)^{2\beta} f(L/\xi) \tag{3.1}$$

as the proportionality function must be dimensionless and has to be independent of $t$. The only variables that can be included are thus $L$ and $\xi$, and to get a dimensionless quantity we can divide those two variables. We neglect here the corrections (1.5).

Using $\xi \sim (-t)^{-\nu}$ (with scaling conditions introduced by equations (1.6) ):

$$\langle m^2 \rangle = \xi^{-\frac{2\beta}{\nu}} f(L/\xi) = L^{-\frac{2\beta}{\nu}} g(L/\xi) = L^{-\frac{2\beta}{\nu}} g\big((-t)^\nu L\big) \tag{3.2}$$

where we just change the function: $g(L/\xi) = (L/\xi)^{-\frac{2\beta}{\nu}} f(L/\xi)$.

When the temperature is equal to the critical temperature, $t \to 0$, so:

$$g\big((-t)^\nu L\big) \to g(0) = const.$$

Therefore, at the critical temperature:

$$\langle m^2 \rangle L^{\frac{2\beta}{\nu}} = const. \quad \text{for } T = T_C$$

Changing the normalized magnetization to the magnetic susceptibility:

$$\langle m^2 \rangle = \frac{\langle M^2 \rangle}{L^2} = \frac{\langle M^2 \rangle}{T} T L^{-2} = \chi\, T L^{-2}$$

Our equation reads:

$$F \equiv \chi\, T L^{\frac{2\beta}{\nu}-2} = const. \quad \text{for } T = T_C \tag{3.3}$$

We can plot this function (3.3), using the critical exponents for 2D Ising model (taken from [22], Table 10.1.1):

$$\nu = 1, \quad \beta = 1/8$$

Our plot of the critical region is shown below.

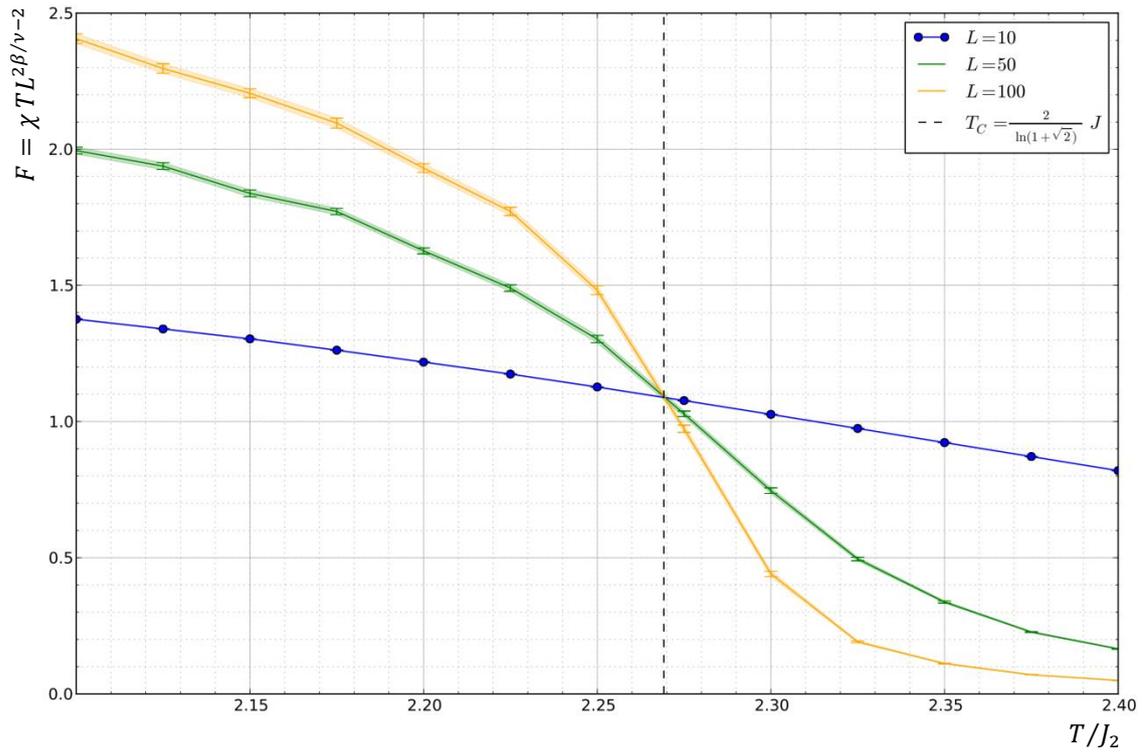

Figure 14. Finding the critical temperature.

To determine the value of the critical temperature, we can zoom the plot on Figure 14 and check the range of temperatures that corresponds to overlaying areas of the errors:

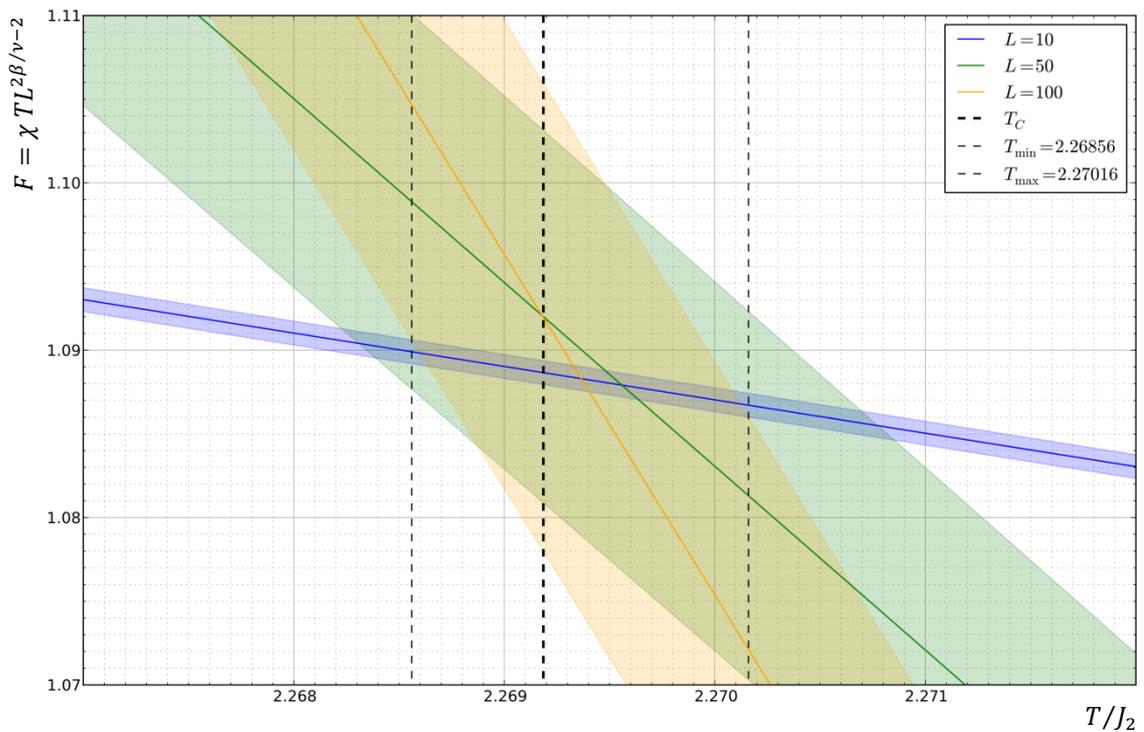

Figure 15. Finding the critical temperature. Dashed lines mark the possible region for value of $T_C$.
It corresponds to the area of all three error areas overlapping.

As we can see the range is $T_C/J \in (2.26856; 2.27016)$, which after proper rounding corresponds to:

$$T_C = (2.2694 \pm 0.0008)\,J \qquad (3.4)$$

We can see that this range includes the analytically determined value of the critical temperature from equation (1.9). Therefore, there is no reason to discard the hypothesis that the simulated value is no different from the expected one. In fact, the value was determined with an outstanding accuracy, with the relative error

$$\varepsilon = \left|\frac{\sigma_{T_C}}{T_C}\right| \approx 0.00035 = 0.035\% \qquad (3.5)$$

being very small value. It is very surprising due to the fact, that when stating the equation (3.1) for order parameter $\langle m^2 \rangle$ we *neglected the corrections-to-scaling* (1.5). This might mean, that while investigating the case of graphene impurities, we can try to use the same process.

# Chapter 4
# Using the worm algorithm
# in the simulation of graphene impurities

## 4.1 Describing graphene impurities and the algorithm to simulate the system

We remember that the graphene impurities prefer to sit on high-symmetry positions (see **Chapter 1.1.2**). In our project we have chosen to simulate the behaviour of the impurities *on top of carbon atom*. It is most common impurity type, due to the most negative binding energy [29]. We then have two sublattices of impurities, corresponding to the sublattices of the graphene lattice itself (see Figure 8c). Thus, we introduce two "colours" to distinguish between the sublattices: some adatoms will be "red" and some will be "blue".

Firstly, let us make some **simplifications**. We consider *magnetic impurities*, that have spin. Moreover, this spin we assume to be classical. In the Ising model we had quantized spin that had only two possible values. In the real life however, spin is a vector that can be angled arbitrary, thus we have infinite number of possibilities. The Ising model with introduced spins as vectors is called *Heisenberg model* and thus we have a slightly changed Hamiltonian:

$$H = -\sum_{i,j} J_{ij} \vec{\sigma}_i \cdot \vec{\sigma}_j \qquad (4.1)$$

Fortunately, because graphene is 2D structure, it is very probable, that if we place the impurity on the lattice, its spin will be $\pm \frac{1}{2}$ and perpendicular to the graphene plane. The reasoning behind it, is that this positioning of spin would maintain as much of symmetry properties as it possibly can.

We move on to the next simplification – we say that the impurities are substantially far from each other. If the impurities distances are much larger than the lattice constant of underlying graphene structure, then we can also say, that we are dealing with the continuous case, where we have some unit area of continuous space and we randomly place $N_{at}$ of impurities on it. The system will be thus described by a *graph*, not a lattice.

Very important base for this model is that the interaction between impurities is *mediated by the conduction electrons*. This should be very similar in nature to the indirect exchange interaction via conducting electrons in solids (called RKKY interaction) [30]. Theoretical predictions [31] show, that those interactions in graphene are changing like $\sim 1/r^3$ and do not oscillate.

Before we write down the Hamiltonian for this model, we have to think about **units of our variables**. In physics most of variables have their unique units. But the quantities that we use in computers are dimensionless. So, when designing the equations that we will use in simulations, we have to bear that in mind. First of all, this time we won't be dealing with the system size $L$ (like in the two-dimensional Ising model simulation), as we are considering unit area. Thus, we have to use the sites density (or adatoms density):

$$\rho = \frac{N_{at}}{L^2} \tag{4.2}$$

Because we have set $L^2 = 1$, then we can see that $N_{at}$ corresponds to $\rho$. The distances will be measured in units of mean distance between impurities. We can define this quantity as:

$$\|R\| = \frac{L}{\sqrt{N_{at}}} = \frac{1}{\sqrt{\rho}}$$

Our dimensionless variable corresponding to distance $R$ will be then:

$$r = \frac{R}{\|R\|} = \rho^{1/2} R$$

Also, all the interaction constants have a dimension of energy, similarly to the Hamiltonian. Thus, when constructing the equation we can use any of their ratios. Our choice here is to present them in the units of interaction constant $J_2$.

Therefore, after all simplifications and including RKKY interactions, we arrive with the Hamiltonian:

$$H = \sum_{i,j} \frac{J_1 - J_2 c_i c_j}{r_{ij}^3} \sigma_i \sigma_j \tag{4.3}$$

where $J_1, J_2$ – interaction constants, $c_i, c_j = \pm 1$ – "colours" of the impurities, $\sigma_i, \sigma_j = \pm 1$ – Ising spins, $r_{ij} = |\vec{R}_{ij}|/\|R\|$ – the normalized distance between site $i$ and $j$. Also, $r_{ij}$ and $c_k$ are "quenched" variables.

The Hamiltonian in a dimensionless form, which we can use in computer simulation is:

$$\frac{H}{J_2} = \sum_{i,j} \frac{J_1/J_2 - c_i c_j}{r_{ij}^3} \sigma_i \sigma_j \tag{4.4}$$

"Colour" of the impurity is just a number that indicates to which sublattice the impurity belongs. If we mark our sublattices as "blue" ($c_i = +1$) and "red" ($c_i = -1$) (cf. Figure 8c), we can see that:

1) If sites $i$ and $j$ are on the same sublattice, then $c_i c_j = 1$ and (4.3) becomes:
$$H = \sum_{i,j} \frac{J_1 - J_2}{r_{ij}^3} \sigma_i \sigma_j \equiv J^{(-)} \sum_{i,j} \frac{1}{r_{ij}^3} \sigma_i \sigma_j$$

2) If sites $i$ and $j$ are on different sublattice, then $c_i c_j = -1$ and (4.3) becomes:
$$H = \sum_{i,j} \frac{J_1 + J_2}{r_{ij}^3} \sigma_i \sigma_j \equiv J^{(+)} \sum_{i,j} \frac{1}{r_{ij}^3} \sigma_i \sigma_j$$

We can see now what the interaction constants mean: $(J_1 - J_2)$ describes the interaction for the same sublattice spins and $(J_1 + J_2)$ is the interaction for the different sublattices spins.

As the reader would have noticed there is no minus sign in front of our Hamiltonian. But we remember, that the worm algorithm was working only if there is this minus sign, i.e. we have ferromagnetic ordering. This would mean, that we are dealing here with the **sign problem**, as it was discussed before in **Chapter 2.2**. However, there are ways to deal with this in some cases. Sometimes we can show that we have ferromagnetic ordering and sometimes we can map antiferromagnetic ordering into ferromagnetic.

In particular, for $J_1 = 0$ equation (4.4) reads:
$$\frac{H}{J_2} = -\sum_{i,j} \frac{1}{r_{ij}^3} \sigma_i \sigma_j c_i c_j = -\sum_{i,j} \frac{1}{r_{ij}^3} s_i s_j, \qquad s_k = \sigma_k c_k = \pm 1$$

We used new variables $s_k$ that behave exactly like the "normal" Ising spins. Thus, we achieved the Hamiltonian for altered Ising model with interaction depending on the distance. This case is **object of our interest** in this project.

Another case is when $J_1 = J_2 = J$. Then the Hamiltonian (4.4) becomes:
$$\frac{H}{J} = \sum_{i,j} \frac{1 - c_i c_j}{r_{ij}^3} \sigma_i \sigma_j = \sum_{i,j} \frac{c_i c_j - 1}{r_{ij}^3} s_i s_j, \qquad s_k = c_k \sigma_k = \pm 1$$

If our considered sites are in the same sublattice, then we get zero interaction. If they are on different sublattices, then $c_i c_j = -1$ and the Hamiltonian term has $-2$ in front, and our sign problem disappears again. This case was calculated before.

In fact, it can be shown (see **Appendix A**), that the range where worm algorithm can be applied is in this model:
$$\frac{J_1}{J_2} \in (-\infty; 1], \qquad J_2 > 0 \tag{4.5}$$

There is an additional problem of very **high disorder** in our simulation. It is important to say that this disorder is not thermal, but "quenched" and is due to $r_{ij}$ and $c_k$. If we place the adatoms on graphene, it is inevitable that they will eventually change their positions (i.e. change $r_{ij}$ and $c_k$). This process is similar to diffusion and is due to non-zero probability that the impurities will tunnel through the energy potential. "Quenched" disorder means that the relaxation of adatoms in a specific position on the lattice corresponds to a much bigger timescale than the relaxation of the spin degrees of freedom. So, we have to calculate the stochastic average of the observable for a given realization of disorder and then take the average over disorder. One can achieve that by making many copies of the system and then average over them. Of course we have to make sure that the seeds of our random generators are different in every copy, so we set them manually.

Usually the standard deviation of the averages given by different copies of the system is essentially large. So we have to include it in our final answer together with the errors given by each system. To do that, we average all errors given by each copy and add it with the standard deviation in the usual way:

$$\sigma_{\text{final}} = \sqrt{\langle \sigma_{\text{copy}} \rangle^2 + \sigma_{\text{population}}^2}$$

Standard deviation of the population is given by:

$$\sigma_{\text{population}} = \sqrt{\frac{1}{N(N-1)} \sum_{i=1}^{N} (x_i - \bar{x})^2}, \qquad \bar{x} = \frac{1}{N} \sum_{i=1}^{N} x_i$$

where $N$ — number of system copies, $x_i$ — value of the observable given by system copy with index $i$.

In our simulation we will be using worm algorithm with *limited configuration space* (see **Chapter 2.3.3**), as its performance is better in the low temperature region.

Now, we can actually sketch the algorithm that we will be using:

1) Place equal number of "blue" and "red" sites randomly on the unit area. Give each site random Ising spin.
2) Choose randomly the site $j_1$ = "ira" = "masha" = $j_2$.
3) Propose an update by choosing randomly the new site $j_3$ and randomly choose if to draw or to erase.
4) Accept $j_3$ as a new "masha", if random real number from region $(0,1)$ is less than the acceptance ratio $R$. We can calculate $R$ using equation (2.9), but with different coupling constant (changed according to (4.4) ):

$$K_{ij} = \beta \cdot \frac{J_1/J_2 - c_i c_j}{r_{ij}^3} \qquad (4.6)$$

5) If $j_1 = j_2$ go back to 2). Otherwise go back to 3).

Having the algorithm we can proceed with our simulation.

## 4.2 Results

Firstly, we set up the simulation:

| Number of impurities $N_{at}$ | Sweeps number | Thermalisation |
|---|---|---|
| 100 | 10 000 000 | 500 000 |
| 1 000 | 1 000 000 | 50 000 |
| (10 000) | (100 000) | (10 000) |
| 10 000 | 1 000 000 | 50 000 |

For $N_{at} = 10\,000$ we firstly used only 100 000 sweeps, because the time of simulation was very long. Unfortunately, the errors that were obtained, were far too large to help us in description of the critical region. So, we increased the sweeps number to one million.

For every $N_{at}$ we used different range and sampling, as in the Ising model simulation:

| $N_{at}$ | $(T/J_2)^2$ | | | |
|---|---|---|---|---|
| | Range | Sampling | Critical region (CR) | Sampling in CR |
| 100 | [1.5; 4.5] | 0.1 | [2.5; 3.5] | 0.05 |
| 1 000 | [1.5; 5.0] | 0.1 | [2.5; 4.5] | 0.05 |
| 10 000 | [1.5; 5.0] | 0.5 | [2.5; 4.5] | 0.10 |

Additional temperature points in the critical region were also simulated for $N_{at} = 10000$, because it was necessary to get good results in determining the critical temperature.

Our goal was to examine the critical region of this model. Unlike in the previous simulation of the Ising model, this one was programmed to only measure value of $\langle M^2/N_{at}^2 \rangle \equiv \langle m^2 \rangle$, the mean magnetization per atom squared. Our result in a form of plot is shown on Figure 16.

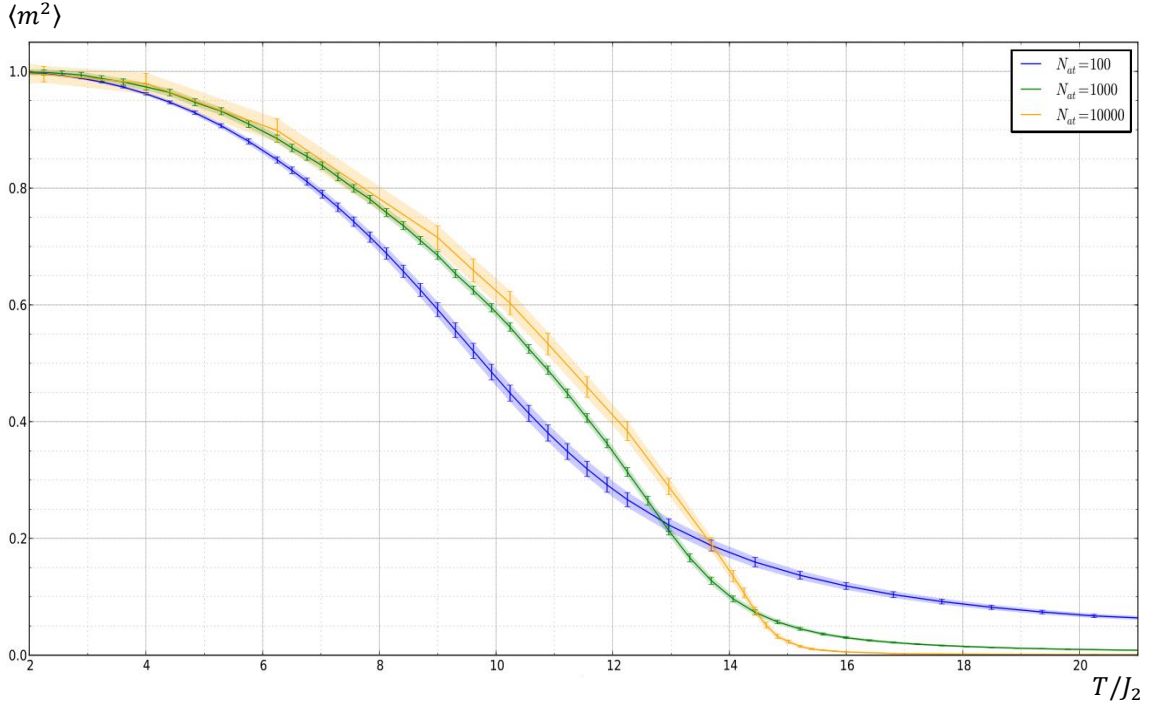

Figure 16. Mean magnetization squared per atom.

## 4.3 Analysis of the results

Using the same reasoning as in **Chapter 3.2**, we can obtain the equation (cf. (3.2) ):

$$\langle m^2 \rangle L^{2\beta/\nu} = f\big((-t)^\nu L\big)$$

What has to be stressed is the fact, that we again *neglect the corrections-to-scaling* (1.5) in the above equation. We don't know, if we can use this simplification, but it is a good start for a "rough" approximation and this procedure was successful in the two-dimensional Ising model simulation.

We must remember, that in this model $L$ has to be expressed by $N_{at}$. To do that, we use equation (4.2) to determine that $L \sim N_{at}^{1/2}$. Therefore, we may write:

$$\langle m^2 \rangle N_{at}^{\beta/\nu} = f\big((-t)^\nu N_{at}^{1/2}\big) \qquad (4.7)$$

We could do the previously used path and say that if $T = T_C$ then $t = 0$ and the function is constant. Unfortunately, this time we don't know what are the critical exponents $\beta$ and $\nu$.

So firstly, we estimate the value of the exponent $\frac{\beta}{\nu}$ and $T_C$ by following what we did in the Ising model. We say that $\langle m^2 \rangle N_{at}^{\beta/\nu}$ must be constant for $T = T_C$. So, we may adjust value of exponent to get area of overlapping errors for all three systems on the plot. Example of this situation is shown on Figure 17.

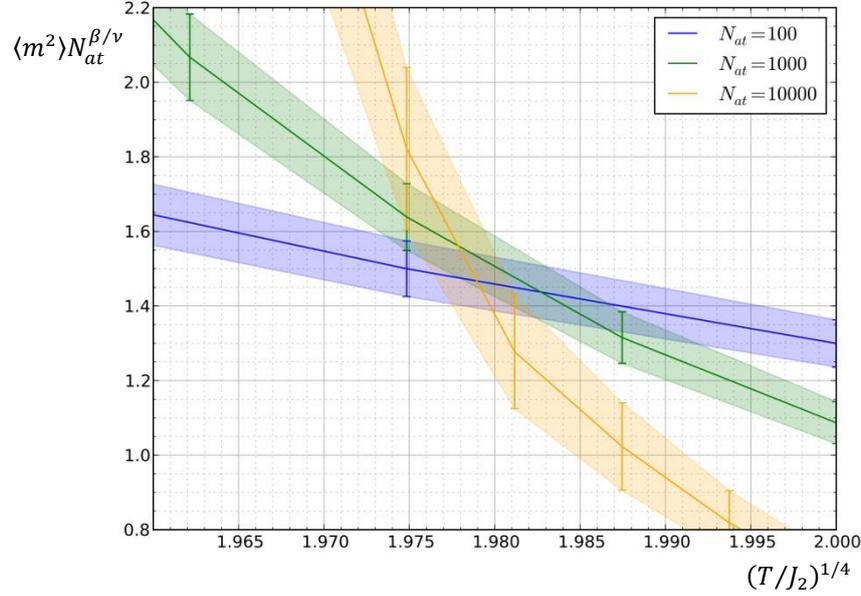

Figure 17. Example of fitting exponent $\beta/\nu$ so that the three error areas are overlapping. Of course some exponents will not give this result, and there will be no space with three overlapping errors.

With sampling for exponent: $0.0001$ we have checked that the minimum and maximum values of exponents that give us this result are $\{0.3914; 0.6122\}$. After appropriate rounding we achieve the estimated value of $\beta/\nu$:

$$\frac{\beta}{\nu} = 0.50 \pm 0.11 \tag{4.8}$$

Estimated value of $T_C$, that corresponds to this region is:

$$T_C = (15.37 \pm 0.41) J_2 \tag{4.9}$$

Furthermore, we want to use function from equation (4.7) to better estimate the exponents. We may change our function a little bit and get:

$$\langle m^2 \rangle N_{at}^{\beta/\nu} = f\left((T - T_C) N_{at}^{1/2\nu}\right) \tag{4.10}$$

We use this equation (4.10) to plot our data. In the critical region the function has to be the same for all systems with different $N_{at}$. We should therefore fit our parameters $\beta$ and $\nu$, so that we would achieve this goal.

So, we use the estimation of exponent $\beta/\nu$ to get an estimation for $\nu$. We use formula (4.10) and for different $\beta/\nu$ we check what are possible values of $\nu$. All three functions for systems with different $N_{at}$ should be overlapping in the critical region.

Best fitting of these functions was shown in Figure 18.

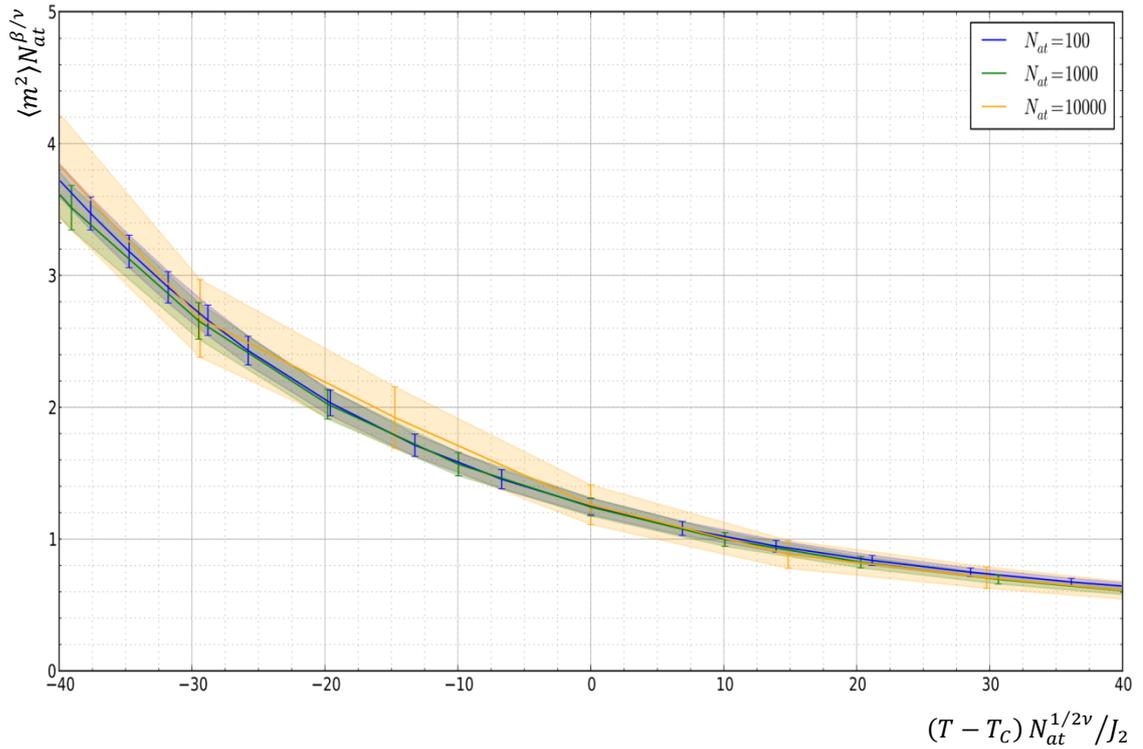

Figure 18. Fitting the functions. Parameters used: $\beta/\nu = 0.48$; $1/2\nu = 0.47$; $T_C = 15.21$.

Our final result for the critical exponents is:

$$\beta \approx 0.51, \qquad \nu \approx 1.1 \qquad (4.11)$$

By comparing those results (4.11) with exponents for various systems (see [22], Table 10.1.1), we can see that our values are very unique and unusual. We remember, that we have neglected the *corrections-to-scaling* (1.5). Many works suggest (example is [32]) that the critical temperature is a quantity easy to determine, while the critical exponents need very high accuracy and are subtle quantities. This means, that while our estimation for critical temperature is accurate while neglecting the corrections, the exponents (4.11) may not be. Thus, we can see, that our next step would be to further investigate the corrections and critical exponents.

# Chapter 5
# Conclusions and outlook

In our project, we have succeeded in simulating the critical region of both Ising model and graphene impurities. The worm algorithm was successfully used in both cases. The simulation of the Ising model gave us remarkably good results – we have showed the phase transition and achieved a very precise value of the critical temperature:

$$T_C = (2.2694 \pm 0.0008)\, J \tag{5.1}$$

Besides, the plots of mean energy, magnetic susceptibility and heat capacity for different sizes of the system shows clearly the phase transition, which is in compliance with the analytical results [19].

Furthermore, we have also showed that the phase transition is possible in the model of graphene impurities and we have attained the critical temperature with the Hamiltonian (4.3) in the case of $J_1 = 0$. The value is:

$$T_C = (15.37 \pm 0.41)\, J_2 \tag{5.2}$$

We have also estimated the critical exponents for this system. With those, we can approximate how magnetization and correlation length behave near the phase transition by using the critical exponents (4.11):

$$m \sim (-t)^{0.51}, \qquad \xi \sim (-t)^{-1.1} \tag{5.3}$$

There was a serious problem of corrections-to-scaling, while we were estimating those values of exponents. We have completely neglected the corrections (1.5), but our results suggest, that they may have a very serious influence on the exponents, since obtained exponents are very unique. Thus, further investigation of the critical exponents *is needed*. We may have to calculate the critical temperature to a very high accuracy (as shown for example in [32]) to get a good values for $\beta$ and $\nu$.

Using our value (5.2) for system $J_1 = 0$ and the value [33]:

$$T_C/J_2 \big|_{J_1 = J_2} = 13.26 \pm 0.28 \tag{5.4}$$

we can now sketch what is the relation between $T_C/J_2$ and $J_1/J_2$.

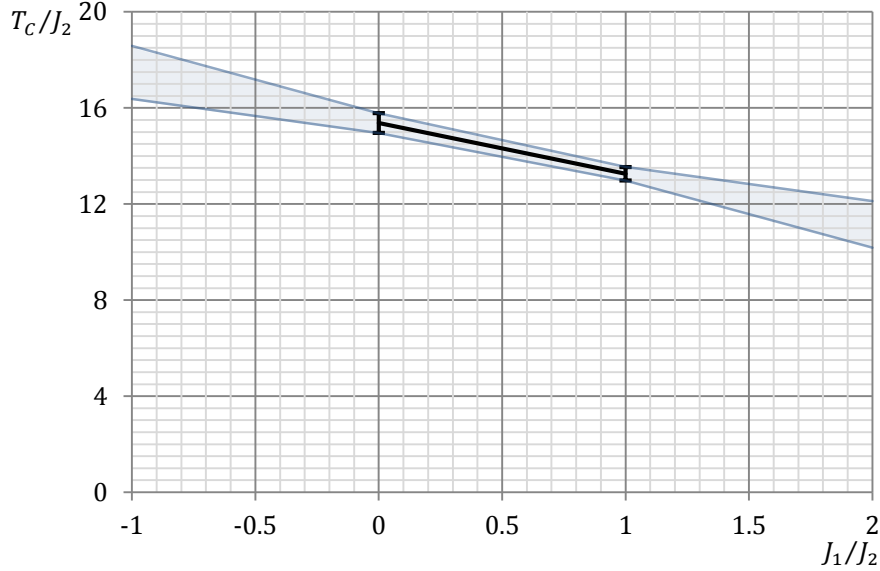

Figure 19. With two points $T_C(0) = (15.37 \pm 0.41)J_2$ and $T_C(1) = (13.26 \pm 0.28)J_2$, we can sketch the dependence $T_C/J_2 = f(J_1/J_2)$. The blue area shows the possible functions, if we have linear relation.

We can see how the Figure 19 shows the possible linear functions, that might describe our system. Further work is necessary to actually determine this dependence, of course. The worm algorithm is not working for case of $J_1/J_2 \in (1; \infty)$, so we will have to use another algorithm for this range. Also, when $J_1/J_2 \to \infty$ we will have the biggest problem, as this is the case of so-called *spin-glass*. We would be dealing then with frustrated spins (i.e. spins will be willing to anti-align with each other – for two spins it is obvious, but if we have three of them, the third will have a problem with directing itself). This is very hard situation to simulate.

Furthermore, values of critical temperature $T_C$ and interaction constant $J_1$ are represented in the units of $J_2$. This latter quantity is yet to be determined.

We can think of improving our modelling of graphene impurities by dropping some simplifications, that we have made in the beginning of **Chapter 4.1**. For example, we can think of a way to introduce spins as vectors, like in Heisenberg model (4.1).

Summarizing, the result of this work is an essential step in describing the phase transition of graphene impurities by the means of computer simulations. We have convincingly shown that the magnetic ordering of adatoms on the graphene lattice is possible and that is leads to the phase transition. The reasonable estimate of critical temperature is achieved, but the further study of critical exponents is necessary. The path that we should follow is outlined and, therefore, we should be able in the end to fully describe, how the critical temperature of impurities on graphene is depending on the interaction constants.

# Appendix A
# Calculating the range
# of the worm algorithm appliance

Let us prove that the range, where we can use worm algorithm in simulating the model of graphene impurities is described by condition $J_1 \leq J_2$.

Our graphene impurities model is described by the Hamiltonian (4.3):

$$H = \sum_{i,j} \frac{J_1 - J_2 c_i c_j}{r_{ij}^3} \sigma_i \sigma_j$$

We know, that the worm algorithm works for the case of Ising model:

$$H = -\sum_{i,j} C_{ij} \sigma_i \sigma_j$$

where $C_{ij}$ is positive interaction constant (or zero in trivial case of no interactions). So, in our case quantity $\frac{J_1 - J_2 c_i c_j}{r_{ij}^3}$ must be never positive. Because $r_{ij} = |\vec{r}_{ij}|$ is always positive, this leads to the constraint:

$$J_1 - J_2 c_i c_j \geq 0$$

It has to be true for all possibilities, so we have to consider cases $c_i c_j = \pm 1$. What has to be stressed, we consider the case, where $J_2 > 0$. So:

| $c_i c_j = +1$ | $c_i c_j = -1$ |
|---|---|
| $J_2 - J_1 \geq 0$ | $-J_2 - J_1 \geq 0$ |
| $J_2 \geq J_1$ | $-J_2 \geq J_1$ |
| $\frac{J_1}{J_2} \leq 1$ | $\frac{J_1}{J_2} \leq -1$ |

This leads to:

$$\frac{J_1}{J_2} \in (-\infty; 1] \cap (-\infty; -1] = (-\infty; -1]$$

But we remember, that we can use the "trick" from **Chapter 4.1**, to introduce pseudo-Ising spins, $s_k = \sigma_k c_k = \pm 1$. If we do that and remember that $(c_i c_j)^2 = 1$, we end up with the Hamiltonian:

$$H = \sum_{i,j} \frac{J_1 c_i c_j - J_2}{r_{ij}^3} s_i s_j$$

Which leads to constraint:

$$J_2 - J_1 c_i c_j \geq 0$$

Continuing the calculations:

| $c_i c_j = +1$ | $c_i c_j = -1$ |
|---|---|
| $J_2 - J_1 \geq 0$ | $J_2 + J_1 \geq 0$ |
| $J_2 \geq J_1$ | $J_2 \geq -J_1$ |
| $\frac{J_1}{J_2} \leq 1$ | $\frac{J_1}{J_2} \geq -1$ |

Which leads to:

$$\frac{J_1}{J_2} \in (-\infty; 1] \cap [-1; \infty) = [-1; 1]$$

Uniting those two ranges gives:

$$\frac{J_1}{J_2} \in (-\infty; +1], \qquad J_2 > 0 \tag{A.1}$$

If $J_2 < 0$, then every time we divide by $J_2$ we have to change the inequality. This leads to:

$$\frac{J_1}{J_2} \in \left( [1; \infty) \cap [-1; \infty) \right) \cup \left( [1; \infty) \cap (-\infty; -1] \right) = [1; \infty) \cup \emptyset$$

$$\frac{J_1}{J_2} \in [+1; \infty), \qquad J_2 < 0 \tag{A.2}$$

We can notice that both ranges (A.1) and (A.2) correspond to inequality:

$$J_1 \leq J_2 \quad \blacksquare$$